\documentclass[12pt]{iopart}
\usepackage{bm}
\usepackage{float}
\expandafter\let\csname equation*\endcsname\relax
\expandafter\let\csname endequation*\endcsname\relax
\usepackage{amsmath}
\usepackage{graphicx}
\usepackage{dcolumn}
\usepackage{cite}
\usepackage{multirow}
\usepackage{footmisc}
\usepackage{amssymb}

\usepackage[dvipsnames]{xcolor}
\usepackage[colorlinks=true,allcolors=RoyalBlue]{hyperref}
\hypersetup{colorlinks=true}
\usepackage{subcaption}
\usepackage{amsfonts}
\usepackage[capitalize]{cleveref}
\usepackage{physics}

\DeclareMathOperator\erfc{Erfc}

\newcommand{\n}{\nonumber}
\usepackage{etoolbox}
\makeatletter
\newrobustcmd{\fixappendix}{%
  \patchcmd{\l@section}{1.5em}{7em}{}{}%
  \patchcmd{\l@subsection}{2.3em}{7em}{}{}%
}
\makeatother

\newcommand{\thetitle}{Dynamics of a tracer trapped in a correlated medium in the presence of a wall
}

\begin{document}
\title[\thetitle]{\thetitle}
\author{Marcin Piotr Pruszczyk$^{1,2}$ and
Andrea Gambassi$^{1,2}$}
\address{$^1$ SISSA --- International School for Advanced Studies, via Bonomea 265, 34136 Trieste, Italy}
\address{$^2$ INFN, via Bonomea 265, 34136 Trieste, Italy}
\ead{mpruszcz@sissa.it}
\begin{abstract}
 We describe the random motion of a particle immersed in a thermally fluctuating medium and harmonically trapped at a certain distance from a wall.  The medium, modeled by a Gaussian field with a tunable correlation length~$\xi$, is linearly coupled to the particle and evolves according to dissipative relaxational dynamics. Dirichlet boundary conditions imposed on the field at the wall give rise to a repulsive fluctuation-induced force  acting on the particle, causing a shift in its average position and a renormalization of the strength of the harmonic trap.
 We describe  the  
 effective overdamped dynamics of the particle, which features a nonlinear memory term depending on the wall-particle separation.
 We show that the two-time correlation function of the particle position features a memory-induced term that depends on the distance of the particle from the wall. At the critical point, this term decays algebraically upon increasing time and  it displays a crossover from the behavior observed in the  bulk to that corresponding to having the particle at the wall.  
 \end{abstract}

\tableofcontents
\markboth{\thetitle}{\thetitle}

\section{Introduction}
\label{sec:intro}
The description of the random motion of mesoscopic particles in media traces back to the seminal works by Einstein \cite{Einstein_1905} and Smoluchowski \cite{Smoluchowski_1906}, and since then, it has remained one of the central themes of statistical physics. In a simple 
fluid, such motion has been successfully described in terms of a linear  Langevin equation for the particle velocity $v$ (see, e.g., Ref.~\cite{Coffey_2004}). In this formulation, the effective interaction between the mesoscopic colloid and the fluid molecules is modeled by a deterministic instantaneous friction $- \gamma v$ proportional to the velocity of the particle, and a random force that has a Gaussian distribution and is temporally uncorrelated.  
This description is valid whenever there is a clear timescale separation, i.e., if the evolution of the degrees of freedom of the medium occurs on timescales substantially shorter than those at which the motion of the particle is described. 

However, when a tracer particle is immersed in a complex medium characterized by macroscopically long relaxation times, the timescale separation underlying the construction of a simple Langevin equation for particle velocity $v$ no longer occurs and the random interaction between the particle and the medium can no longer be described by an uncorrelated Gaussian noise. 
Accordingly, owing to the fluctuation-dissipation theorem in equilibrium \cite{Kubo_book}, 
the deterministic friction experienced by the particle at time $t$ is no longer instantaneous, i.e., it does not depend solely on $v(t)$. 
The presence of hydrodynamic memory \cite{Franosch2011} or viscoelasticity in the 
medium \cite{Gomez-Solano_2015, Ginot_2025} provides concrete examples. In these cases, in fact, the effect of the medium on the particle dynamics is described by a retarded response $\Gamma(t)$, which determines 
the effective friction force as $-\int^t \rmd  t' \, \Gamma(t-t')v(t')$. 
The kernel $\Gamma(t)$ can be experimentally inferred from studying the spectrum of the equilibrium fluctuations of the position \cite{Mason_1995} 
of the probe. 
In practice,  $\Gamma(t)$ can be determined in microrheology based on the measurement of the viscoelastic shear modulus of the medium, which is related to it and which is obtained from the observation of the motion of the probe \cite{Mason_1995}. This method has been widely used in experimental investigations of the statistical properties of soft matter in recent years, see, e.g., Refs.~\cite{Wirtz_2009,Daldrop_2017,Muller_2020}.

The effective description of the dynamics of the probe in terms of a linear memory kernel, though applicable and accurate in many contexts, turns out to be generically insufficient for describing colloidal particles subject to optical trapping. In fact, as shown both numerically and experimentally \cite{Daldrop_2017, Muller_2020},  the resulting memory kernel $\Gamma(t)$ turns out to depend on the details of the confinement, in particular on the trapping forces acting on the particles. This means that $\Gamma$ loses the desirable feature of being determined solely by the interaction between the particle and the medium. 
This observation can be explained by the fact that the effective dynamics of the probe is actually  governed by a nonlinear evolution equation \cite{Basu_2022}.  In the simple model for which this confinement-dependent memory was derived, the medium is described by a background solvent and a fluctuating Gaussian field $\phi$. This correlated field, characterized by a tunable spatial correlation length~$\xi$,  follows relaxational and locally conserved dynamics, i.e., the so-called (Gaussian) model B dynamics \cite{Halperin_1977}. 
The motion of the colloid is described by an overdamped Langevin equation with instantaneous friction due to the interaction with the background solvent. Additionally, the probe is linearly coupled to $\phi$ \cite{Demery_2010, Demery_2010_2, Demery_2011, Dean_2011}.  
Both the field $\phi$ and the particle are in contact with the solvent, which acts as a thermal bath. This model and its variants were employed to investigate the dynamics of particles that diffuse freely or are dragged \cite{Demery_2011, Demery_2019, Demery_2014} in the bulk or under spatial confinement \cite{Gross_2021, Venturelli_Gross_2022}, and were successful in describing a variety of phenomena (both in and out of equilibrium) due to the presence of correlations in the medium \cite{ Venturelli_2022, Venturelli_Gross_2022, Venturelli_2022_2parts, Venturelli_2024,  Demery_Gabassi_2023, Venturelli_2023, Pruszczyk_2025}.

The possibility of tuning the correlation length $\xi$ and, consequently, the relaxation time of the fluctuations of the field is a property of media in the vicinity of the critical point at which they undergo a second-order phase transition. 
A concrete example is provided by any binary liquid mixture in the vicinity of its demixing point at which $\xi$ diverges \cite{Huang_book}.
In these cases, however, the description of the dynamics of the media requires more refined models (see, e.g., Refs.~\cite{Halperin_1977, Tauber_2014}) than those considered here.  
Due to the presence of fluctuations with long-range correlations, bodies immersed in these media (imposing boundary conditions on such fluctuations at their surfaces) experience a fluctuation-induced force, known as the  critical Casimir force \cite{Kardar_1999, Krech_book, Brankov_book, Gambassi_2009, gambassi_critical_2024},
which is a thermodynamic analogue of the Casimir force observed in quantum electrodynamics \cite{Casimir_1948}. Since the first theoretical prediction of the occurrence of critical Casimir forces by Fisher and de Gennes \cite{Fisher_1978}, these forces have been investigated theoretically in a variety of systems and within various approaches, including lattice spin systems \cite{Evans_1994, Dantchev_2004}, quantum gases \cite{Martin_2006, Napiorkowski_2013}, and field-theoretical methods \cite{krech_diet_1992, Vasilyev_2013}. Moreover, they were experimentally measured, first indirectly in the film geometry \cite{Garcia_1999, Garcia_2002,  Fukuto_2005, Ganshin_2006, Rafai_2007}, 
and then directly, acting on a colloidal sphere close to a flat surface \cite{Hertlein_2008}. 
These forces emerge also by confining the long-range fluctuations occurring in non-equilibrium conditions \cite{Ortiz_de_Zarate_book}.
While fluctuation-induced forces have been widely investigated at equilibrium,  their dynamics or occurrence out of equilibrium are still largely unexplored. 
In this direction, these forces have been investigated in the film geometry with fixed surfaces after a sudden temperature
change \cite{Gambassi_2008_crit_relax, Rohwer_2017, Gross_2019}, after the action of an external force  \cite{Gambassi_2006_dyn}, or in non-equilibrium steady states  \cite{Cattuto_2006, Brito_2007,Fava_2024}. 
The case in which the confining surfaces are allowed to move under the effect of the fluctuation-induced forces was also investigated in some instances, both theoretically 
\cite{Furukawa_2013, Venturelli_2022_2parts, Gross_2021, Venturelli_Gross_2022} and experimentally 
 \cite{Martinez_Entropy_2017} in media close to a second-order phase transition.

The goal of the present study is to describe the interplay between the emergent memory in the effective dynamics of a tracer immersed in a correlated medium (see, e.g., Ref.~\cite{Basu_2022}) and the fluctuation-induced force acting on the tracer.  
First, in \cref{sec_model}, we introduce the model  
in which the medium fills the space on one side of a $(d-1)$-dimensional flat surface (wall) at which it satisfies Dirichlet boundary conditions. The particle  is coupled linearly to the field, and it is kept away from the wall by a harmonic trap with its center at a distance $X_0$ from it. 
Both the particle and the field are in contact with an equilibrium thermal bath at temperature $T$. The particle evolves according to an overdamped Langevin dynamics, and the field follows relaxational dissipative dynamics \cite{Halperin_1977}, which satisfy the detailed balance condition.
The equilibrium probability distribution function of the particle is investigated in  \cref{sec_eq_pdf}, and it is shown to feature an effective potential -- representing the average effect of the fluctuation-induced force on the probe -- which is repulsive and thus pushes the particle away from the wall. 
The effective wall-particle interaction at the critical point decays as  $\sim X_1^{-(d-1)}$ with increasing wall-particle separation $X_1$, where $d$ is the spatial dimensionality of the system.
This behavior actually characterizes the critical Casimir force (which, strictly speaking, emerges when \emph{all} the involved surfaces --- and not solely the wall --- impose boundary conditions on the field) within the so-called small sphere approximation \cite{Eisenriegler_1995}. 
In the case of a one-dimensional system $d=1$ and in the limit of a point-like particle,  we describe the resulting shift of the average particle position and the suppression of thermal fluctuations quantified by its variance, due to an effective renormalization of the trap strength. 

In \cref{sec_eff_dyn}, 
we formulate an  exact effective nonlinear and non-Markovian dynamics of the tracer, which is not invariant under spatial translations due to the presence of the wall. We show that a relation connecting the correlations of the additional field-induced noise with the effective nonlinear friction acting on the tracer ensures that the dynamics is invariant under time reversal.
In \cref{sec_pert_exp}, we perform a perturbative expansion in the field-particle coupling constant. In $d=1$ and in the limit of a point-like particle, we evaluate the lowest-order correction to the correlation function of the particle position. It exhibits two terms: one corresponding to the renormalization of the trap strength in the Ornstein-Uhlenbeck process, and the other stemming from memory effects present in the system. 
We note that, at the critical point, the latter features an algebraic behavior $\sim t^{-1/2}$ at long times $t$, and a crossover from an asymptotic curve characterizing the particle in the bulk to the one describing the particle trapped exactly at the wall. This stems from the ``propagation'' of correlations in the system, similarly to what was previously reported in  \cite{Furukawa_2013, Venturelli_2022_2parts}. 
We also describe the power spectral density (PSD) of the process and discuss whether the presence of the wall enhances the net correlations of the particle motion, characterized by the PSD at zero frequency. Finally, we summarize our results and formulate an outlook in \cref{sec_conclusions}.

\section{The model}
\label{sec_model}

We consider a system consisting of a tracer particle, modeling a colloid, located at position $\mathbf{X}$, and a correlated medium $\phi$, with the effective Hamiltonian~\cite{Venturelli_2022}
\begin{equation}
    \mathcal{H}[\phi,\mathbf{X}] = \mathcal{H}_\phi[\phi]  +
    \mathcal{H}_{\mathrm{int}}[\phi,\mathbf{X}] + \mathcal{U}(\mathbf{X}).
    \label{eq:hamiltonian_rec}
\end{equation}
The correlated medium is modeled by a fluctuating scalar field $\phi(\mathbf{x}, t) \in \mathbb{R}$ characterized by the quadratic Hamiltonian~\cite{Halperin_1977}
\begin{equation}
        \mathcal{H}_\phi[\phi]= \int_{\Lambda} \mathrm{d}^d{\mathbf{x}}\left[ \frac{1}{2}(\nabla\phi)^2+\frac{1}{2}r\phi^2\right].
        \label{eq:gaussian_hamiltonian_rec}
\end{equation}
Here, we consider the system in a semi-infinite geometry, i.e., the field is defined for $\mathbf{x} = (x_1, x_2, \ldots, x_d) \in \Lambda = \mathbb{R_+} \times \mathbb{R}^{d-1}$, that is, for $x_1\ge 0$. Physically, this corresponds to having an impenetrable $(d-1)$-dimensional flat wall placed at $x_1=0$ that confines the fluctuating field to the space with $x_1>0$. The interaction between the field and the wall is assumed to be effectively modeled by imposing 
Dirichlet boundary conditions (BCs)
 \begin{equation}
     \phi(\mathbf{x}){\vert_{x_1 = 0}} = 0.
     \label{BCs}
 \end{equation}
This and other boundary conditions can be implemented directly in the effective Hamiltonian of the field by introducing the surface contribution \cite{gambassi_critical_2024}
 \begin{equation}
     \int\mathrm{d}^{d-1}\mathbf{x}_{\parallel} \left\{ \frac{c}{2}\phi^2(x_1 = 0, \mathbf{x}_{\parallel}) - h\phi(x_1 = 0, \mathbf{x}_{\parallel}) \right\},
     \label{surf_Ham}
 \end{equation}
where $\mathbf{x} = \left(x_1, \mathbf{x}_{\parallel} \right)$. Here, the surface coupling constant $c$ is the so-called surface enhancement and $h$ is the so-called surface field \cite{gambassi_critical_2024}. Importantly, in the absence of symmetry-breaking terms at the boundaries (i.e., for $h = 0$), and in the limit $c \to +\infty$, the Dirichlet boundary conditions mentioned above are recovered.    
In passing, we note that in the film geometry $[0,L] \times \mathbb{R}^{d-1}$, in which the field interacts with two parallel walls placed at $x_1 = 0$ and $x_1 = L$ via (possibly different) BCs (e.g., Dirichlet, Neumann, periodic or antiperiodic), fluctuation-mediated critical Casimir forces acting on the walls are observed \cite{Kardar_1999, Gambassi_2009,  Napiorkowski_2011, Napiorkowski_2013, Schmidt_2023, gambassi_critical_2024}.  Importantly, the choice of BCs plays a crucial role in determining the properties of the critical Casimir forces at equilibrium 
     \cite{Dantchev_2023, Gambassi_2009, gambassi_critical_2024}. 
     In fact, not only does the strength of the fluctuation-induced force depend on such a choice (although the dependence is largely universal), but also its attractive or repulsive character.
The parameter $r>0$ in \cref{eq:gaussian_hamiltonian_rec} determines the spatial extension $\xi$ of the correlations of the fluctuations of the field. Actually, it  is related to the correlation length $\xi$ via $\xi= r^{-\nu}$ with the critical exponent $\nu = 1/2$, see Ref.~\cite{Zinn_Justin_small}. 
In this respect, the thermally fluctuating medium  can be interpreted as the order parameter of a second-order phase transition occurring at $r = 0$. For instance, 
it can be identified with the relative concentration of the two components of a binary liquid mixture close to its demixing point. 

The tracer modeling a colloidal particle \cite{Dean_2011, Basu_2022, Venturelli_2023, Venturelli_2024, Venturelli_2022_2parts, Venturelli_2022, Venturelli_Gross_2022, Demery_Gabassi_2023, Pruszczyk_2025} is described by its position $\mathbf{X}=(X_1,X_2, \ldots, X_d)$ in the semi-infinite $d$-dimensional space $\Lambda$.  It is subject to a harmonic potential
\begin{equation}
    \mathcal{U}(\mathbf{X})=  \frac{\kappa}{2}\left( \mathbf{X} - \mathbf{X}_0 \right)^2
    \label{def:Uk_rec}
\end{equation}
with stiffness $\kappa$ and a minimum at position $\mathbf{X}_0 = \hat{\mathbf{e}}_1 X_0$, where $\hat{\mathbf{e}}_1$ is the unit vector along the first spatial direction, and $X_0>0$ is the distance between the wall and the trap minimum.
Additionally, the particle is coupled linearly to the field in \cref{eq:hamiltonian_rec} by
\begin{equation}
    \mathcal{H}_\mathrm{int}[\phi,\mathbf{X}] = -\lambda \int_{\Lambda} \mathrm{d}^d  \mathbf{x}\,\phi(\mathbf{x})V\left(\mathbf{x} -\mathbf{X}\right),
    \label{eq:Hint}
\end{equation}
where the interaction kernel $V(\mathbf{x})$ actually models the shape of the particle and is normalized by requiring that its integral over $\mathbb{R}^d$ is one. Accordingly, the strength of the interaction between the field and the tracer is determined solely by the value of the parameter $\lambda$. 
In passing, we note that due to the presence of the wall at $x_1=0$, this coupling is not invariant under translations in space. This is in contrast with the cases investigated in Refs.~\cite{Basu_2022, Venturelli_2023, Venturelli_2024, Venturelli_2022_2parts, Venturelli_2022,  Demery_Gabassi_2023, Pruszczyk_2025}, in which the spatial integral in \cref{eq:Hint} was over $\mathbb{R}^d$ instead of $\Lambda$. 
The coupling between the field $\phi$ and the particle is chosen to be linear for two reasons: 
\begin{itemize}
\item[(i)] it breaks the $\phi \mapsto -\phi$ symmetry of the Hamiltonian of the field \eqref{eq:gaussian_hamiltonian_rec}, and thus, for positive $\lambda V(\mathbf{x})$, it favors configurations of $\phi$ such that it is enhanced in the vicinity of the particle.
When $\phi$ is interpreted as the order parameter of a demixing transition, the enhancement of the field around the particle models a preferential adsorption of one of the two components by the surface of the particle;
\item[(ii)] it lends itself to an exact solution of the dynamics of the field in terms of the particle coordinate $\mathbf{X}(t)$, which, in turn, enables the formulation of an exact effective dynamics of the position of the particle, as discussed further below. 
\end{itemize}
It is worth mentioning that a linear coupling similar to Eq.~\eqref{eq:Hint} emerges  between the bath density-fluctuation field and the position of the tracer particle when describing a tracer in a bath of dense soft colloids via the linearization of the Dean equation, see Refs.~\cite{Kawasaki_1994, Dean_1996, Demery_2014, Demery_2019}.
Additionally, $V(\mathbf{x})$ is assumed to be spherically symmetric and characterized by a single lengthscale, i.e.,  $V(\mathbf{x}) =   \hat{V}(|\mathbf{x}|/R)$, where $R$ plays the role of the effective radius of the particle.
The Gaussian interaction kernel 
\begin{equation}
            V^{\rm G}(\mathbf{x}) = \frac{1}{(2 \pi R^2)^{d/2}}\exp\left({-\frac{\mathbf{x}^2}{2 R^2}}\right)
            \label{eq:Gauss_ker}
        \end{equation}
and the exponential interaction kernel
        \begin{equation}
            V^{\rm e}(\mathbf{x}) = \frac{1}{2 \pi^{d/2}  R^d}\frac{\Gamma(d/2)}{\Gamma(d)}\exp\left(-|\mathbf{x}|/R\right)
            \label{eq:exp_ker}
        \end{equation}
may serve as examples. We note that in the limit $R \to 0$ of a point-like particle, both $V^{\rm G}(\mathbf{x})$ and $V^{\rm e}(\mathbf{x})$ satisfy
        \begin{equation}
            V^{\rm G/e}(\mathbf{x}) \xrightarrow[R \to 0]{} \delta^{(d)}(\mathbf{x}).
        \label{eq:kernel_point_like_lim}
        \end{equation}
We note that in the effective description of the interaction between the particle and the field introduced in Eq.~\eqref{eq:Hint}, the field
does not actually vanish in the interior of the interaction kernel $V(\mathbf{x})$, which we identify with the ``particle''. 
Note that after choosing the functional form of the interaction kernel, the coupling is described in terms of two effective parameters: (i) the coupling constant $\lambda$, and (ii) the effective particle radius~$R$.  

For future convenience, we recall here that the physical dimensions $[\phi]$ and $[\lambda]$ of the field and the coupling, respectively, which follow from \cref{eq:hamiltonian_rec}, are given by $[\phi] =  \mathcal{E}^{1/2} \mathcal{L}^{1 - d/2}$
and $[\lambda] = \mathcal{E}^{1/2} \mathcal{L}^{d/2 - 1}$
in units of energy $\mathcal{E}$ and length $\mathcal{L}$.

\subsection{Dynamics}

We describe the dynamics of the position $\bm{X}$ of the particle  in terms of the overdamped Langevin equation 
\begin{equation}
        \gamma_0 \dot{\mathbf{X}}= -\bm{\nabla}_{\mathbf{X}} \mathcal{H}   + \bm{\eta} ,\label{eq:particle_rec}
\end{equation}
where $\gamma_0$ is the drag coefficient of the tracer, and $\bm{\eta}$ is a stochastic Gaussian noise with a vanishing mean and variance given by (for simplicity, the Boltzmann constant is set to unity here and henceforth, i.e., $k_B \equiv 1$)
\begin{equation}
    \left \langle\eta_{i}(t) \eta_{j}(t') \right \rangle = 2 \gamma_0 T \delta_{ij}\delta(t-t'),
    \label{eq:part_noise}
\end{equation}
where $T$ is the temperature of the equilibrium heat bath with which the particle is in contact. 
The evolution of the field $\phi$ is governed by the relaxational dynamics~\cite{Tauber_2014}
\begin{equation}
\partial_t\phi(\mathbf{x},t)= -D\frac{\delta{\mathcal{H}[\phi, \mathbf{X}]}}{\delta \phi(\mathbf{x},t)} + \zeta(\mathbf{x},t),
\label{eq:field_rec} 
\end{equation}
where $D$ is the mobility of the field, and $\zeta(\mathbf{x},t)$ is the noise discussed further below. We note that \cref{eq:field_rec} cannot be cast in the form of a continuity equation, and the field $\phi$ is not locally conserved. The dissipative dynamics considered here corresponds to the so-called model~A, according to the classification of Ref.~\cite{Halperin_1977}.  
In the current analysis, the self-interaction term $\propto \phi^4$ present in the original formulation of this model is neglected, and we focus on its Gaussian approximation. In previous studies \cite{Basu_2022, Demery_Gabassi_2023, Venturelli_2022, Venturelli_2022_2parts, Venturelli_2023, Venturelli_2024, Pruszczyk_2025} 
in which the particle and the field were in the bulk (i.e., with $X_0 \to \infty$),  the field $\phi$ was locally conserved in the dynamics according to the so-called model~B \cite{Halperin_1977}. 
 In the bulk, this amounts to having $D \mapsto -D \nabla^2$ in \cref{eq:field_rec}, and modifying the correlations of the noise described below according to $\left \langle\zeta(\mathbf{x},t)\zeta(\mathbf{x}',t')\right \rangle \mapsto - \nabla^2 \left \langle\zeta(\mathbf{x},t)\zeta(\mathbf{x}',t')\right \rangle$.
In such a case, \cref{eq:field_rec} can be cast in the form $\partial_t \phi = - \bm{\nabla} \cdot \mathbf{J}(\mathbf{x}, t)$, where $\mathbf{J}$ is the corresponding current.
However, in the presence of boundaries, the relationship above between the two models no longer applies. In fact, as mentioned in Ref.~\cite{Venturelli_Gross_2022} and discussed in Refs.~\cite{Diehl_1992, Gross_2019}, the simple Dirichlet BCs in Eq.~\eqref{BCs} imposed on the field at the wall in model A are incompatible with the conserved dynamics because they lead to a non-zero flux of the field across the boundaries, which causes the field to be locally conserved away from the wall, but not globally. Imposing global conservation on the field  in such a case is technically involved \cite{Gross_2018}. In order to gain analytical insight into the case of Dirichlet boundary conditions, we thus assume $\phi$ to follow dissipative dynamics. 

Both the field and the particle are in contact with the same heat reservoir which is characterized by the temperature $T$. Consequently, the noise 
$\zeta(\mathbf{x}, t)$  acting on the field $\phi(\mathbf{x}, t)$  is a Gaussian variable with vanishing mean and variance
\begin{equation}
    \left \langle\zeta(\mathbf{x},t)\zeta(\mathbf{x}',t')\right \rangle= 2DT  \delta^{(d)}(\mathbf{x}-\mathbf{x}')\delta(t-t').
    \label{eq:field_noise_rec}
\end{equation}
Note that the correlations of the noise given by Eqs.~\eqref{eq:part_noise} and \eqref{eq:field_noise_rec} satisfy the fluctuation-dissipation relation. 
As a consequence, the joint dynamics of the particle and the field is invariant under time reversal. Accordingly, the joint probability distribution function of the position of the particle and the field configuration in the resulting equilibrium steady state is given by the Boltzmann distribution $\propto \exp\left({-\mathcal{H}[\phi, \mathbf{X}]}/T\right)$, where $\mathcal{H}$ is the Hamiltonian of the system introduced in Eq.~\eqref{eq:hamiltonian_rec}.

    For a system described by the Hamiltonian in Eqs.~\eqref{eq:hamiltonian_rec},~\eqref{eq:gaussian_hamiltonian_rec},~\eqref{def:Uk_rec}, and \eqref{eq:Hint}, the equations of motion \eqref{eq:particle_rec} and \eqref{eq:field_rec} read
    \begin{align}
        \partial_t\phi(\mathbf{x},t) = -D \left[(r - \nabla^2)\phi(\mathbf{x},t) -   \lambda V\left(\mathbf{x} - \mathbf{X}\right) \right] + \zeta(\mathbf{x},t), 
                    \label{dyn_both1} \\ 
	              \gamma_0 \dot{\mathbf{X}}(t) = -\kappa \left[ \mathbf{X}(t) - \mathbf{X}_0 \right] - \lambda \int_{\Lambda} \mathrm{d}^d\mathbf{x}\,  \phi(\mathbf{x})\bm{\nabla}V(\mathbf{x} - \mathbf{X}(t)) + \bm{\eta}(t).
        \label{dyn_both2}	
        \end{align}
As in the absence of boundaries \cite{Basu_2022}, 
these equations of motion and the imposed BC turn out to be invariant under the transformation 
    \begin{equation}
        \left\{\phi, \lambda \right\} \mapsto \left\{-\phi, -\lambda \right\},
        \label{parity}
    \end{equation}
which will prove itself useful in \cref{sec_pert_exp}. 
    
\section{Equilibrium Probability Distribution Function}
    \label{sec_eq_pdf}
    
In view of the fact that the stochastic dynamics described by Eqs.~\eqref{eq:particle_rec}~and~\eqref{eq:field_rec}  with the noise terms in Eqs.~\eqref{eq:part_noise}~and~\eqref{eq:field_noise_rec} satisfy the fluctuation-dissipation relation, the resulting steady state is the equilibrium one \cite{Coffey_2004}, and the joint probability distribution function of the particle position and the field configuration is given by the Gibbs-Boltzmann distribution
            \begin{equation}
            P(\mathbf{X}, \phi(\mathbf{x})) = \mathcal{N}^{-1}  \exp\left\{-\frac{1}{T}\mathcal{H}[\phi, \mathbf{X}]\right\},
            \label{Gibbs_full}
            \end{equation}
where $\mathcal{N}$ is a normalization factor and $\mathcal{H}[ \phi, \mathbf{X}]$ is the Hamiltonian introduced in \cref{eq:hamiltonian_rec}. 
Note that when the particle is decoupled from the field, i.e., when $\lambda=0$, the probability density $P(\mathbf{X}, \phi(\mathbf{x}))$ factorizes. Accordingly, neglecting the possible effects that the presence of the wall may have on the particle (i.e., assuming $X_0 \to \infty$) yields a Gaussian distribution for the particle position, which is centered around $\mathbf{X_0}$ and has  
           \begin{equation}
               \left [ \langle X_i X_j \rangle -\langle X_i\rangle \langle X_j \rangle \right]_{\lambda=0,  X_0\to \infty}  = \delta_{i,j} \frac{T}{\kappa},
               \label{eq:variance_decoupled}
           \end{equation}
where the subscript $\lambda=0$,  $X_0\to \infty$ denotes that the averages are calculated for a particle in the bulk, decoupled from the field.  
This leads to a natural definition of the thermal lengthscale  
           \begin{equation}
               \ell_T = \sqrt{T/\kappa},
               \label{eq:lt_def}
           \end{equation}
which describes the spatial spreading of the probability density and thus quantifies the spatial extension of the fluctuations of the position of the particle. 
In this work, we are primarily interested in the effects of the coupling of the particle to the fluctuating field satisfying the boundary conditions in Eq.~\eqref{BCs}  and not in those occurring when the particle approaches the wall. Accordingly, henceforth, we assume that 
            \begin{align}
            &X_0/\ell_T \gg 1, 
            \label{eq:X0_gg_lT}
            \\  &X_0/R \gg 1,
            \label{eq:X0_gg_R}
            \end{align}
unless stated otherwise. 
            
When the particle and the field are coupled, i.e., for $\lambda \neq 0$, they are no longer statistically independent, and the marginal probability distribution $P(\mathbf{X})$ of the particle  position~$\mathbf{X}$ is obtained by integrating out the field degrees of freedom in the joint probability distribution~\eqref{Gibbs_full}, i.e., it is given by
		\begin{equation}
			P(\mathbf{X}) = \mathcal{N}^{-1} \int \mathcal{D'}\phi \exp\left\{-\frac{1}{T}\mathcal{H}[ \phi, \mathbf{X}]\right\},
            \label{eq:pdf_marg_def}
            \end{equation}
where  $\mathcal{D'}\phi$ denotes the functional integration over the configurations of the field satisfying the constraint in Eq.~\eqref{BCs}.
The presence of the wall and the corresponding Dirichlet boundary conditions imposed on the field $\phi(\mathbf{x})$ at $x_1 = 0$ break the spatial translational invariance of the coupling between the colloid and the medium. Accordingly, the marginal distribution $P(\mathbf{X})$ \emph{is not} determined solely by the external harmonic trapping in Eqs.~\eqref{eq:hamiltonian_rec} and \eqref{def:Uk_rec} (as it occurs in the absence of the wall, see Ref.~\cite{Basu_2022}), but also by the distance $X_1$ between the center of the probe and the wall. The resulting distribution $P(\mathbf{X})$ can be conveniently parameterized as 
            \begin{equation}
                P(\mathbf{X}) = \mathcal{N}_X^{-1} \exp\left\{  -\frac{1}{T} \left[\frac{\kappa (\mathbf{X} - \mathbf{X}_0)^2}{2} + V_{\mathrm{eff}}(X_1) \right] \right\},
                \label{pdf_eq}
            \end{equation}
where $\mathcal{N}_X$ is the normalization factor and $V_{\mathrm{eff}}(X_1)$ is 
the effective potential given by
\begin{align}
			V_{\mathrm{eff}}(X_1) = &-\frac{\lambda^2}{2} \int_{\Lambda} \mathrm{d}^d\mathbf{x}\int_{\Lambda} \mathrm{d}^d\mathbf{y}\, V(\mathbf{x} - \mathbf{X}) 
            C^D_{\phi,\rm{st}}(\mathbf{x},\mathbf{y})
            V(\mathbf{y} - \mathbf{X}),
   \label{V_eff}
		\end{align}
in terms of the potential $V$ and the static (i.e., equal-time) correlation function $C^D_{\phi,\rm{st}}(\mathbf{x},\mathbf{x}') = \langle \phi(\mathbf{x}) \phi(\mathbf{x}') \rangle$ of the field in the presence of the wall, which satisfies Dirichlet boundary conditions, i.e., it vanishes as $\mathbf{x}$ or $\mathbf{x}'$ approaches the wall. This correlation function can be expressed as 
\begin{equation}
C^D_{\phi,\rm{st}}(\mathbf{x},\mathbf{y}) = C_{\phi,\rm{st}}\left(  \mathbf{x}-  \mathbf{y}\right) - C_{\phi,\rm{st}}\left(  \mathbf{x}-  \mathbf{y}_R\right),
\label{Dpropagators}
\end{equation}
where $\mathbf{y}_R \equiv - y_1 \mathbf{\hat{e}}_1 + \mathbf{y}_{\parallel}$ is the vector obtained by reflecting $\mathbf{y} = y_1 \mathbf{\hat{e}}_1 + \mathbf{y}_{\parallel}$ with respect to the wall, i.e., its ``image''. In the expression above, $C_{\phi,\rm{st}}(\mathbf{x})$ is the correlation function of the Gaussian field in the bulk, given by \cite{Zinn_Justin_small} 
        \begin{equation}
			C_{\phi,\rm{st}}(\mathbf{x}) = \int\frac{\mathrm{d}^d \mathbf{k}}{(2 \pi)^d} \frac{e^{-i\mathbf{k}\cdot \mathbf{x}}}{k^2 + r} = \begin{cases}
			    \frac{\xi}{2} {\rm e}^{-|x|/\xi} & \mbox{for\ } d=1,  \\ 
                \frac{1}{2\pi}  K_0\left(|\mathbf{x} |/\xi\right) & \mbox{for\ } d=2, \\ 
         \frac{1}{4 \pi|\mathbf{x}|}{{\rm e}^{-|\mathbf{x}|/\xi}} & \mbox{for\ } d=3, 
         \label{propagators}
			\end{cases}
		\end{equation}
where $K_0$ is the modified Bessel function of the second type. 
The derivations of these expressions are provided in~\ref{per}. 
Note that the expression reported in \cref{pdf_eq} holds for arbitrary values of the coupling $\lambda$ describing the field-particle interaction (see \cref{eq:Hint}). 
Since the particle does not interact directly with the wall,  the effective potential $V_{\mathrm{eff}}(X_1)$ describes the interaction mediated by the fluctuating field.
Accordingly, we call the fluctuation-induced force 
\begin{equation}
          F_{\mathrm{C}}\left(X_1\right) = -\frac{\mathrm{d}}{\mathrm{d} X_1}V_{\mathrm{eff}}(X_1)
          \label{Cas_def}
\end{equation}
associated with this potential 
$V_{\mathrm{eff}}$ the Casimir-like force \cite{gambassi_critical_2024}.   
We note that the term \textit{Casimir force} usually refers to the effective interaction between two bodies that impose boundary conditions on the fluctuations of a medium \cite{Gambassi_2009} at their surfaces.  
Here, instead, the boundary conditions are imposed on the field only on the wall, while the coupling between the field and the particle occurs via an interaction potential $V(\mathbf{x})$, according to \cref{eq:Hint}. 
However, as in a previous study \cite{Gross_2021}, we refer to the force in this case as the (critical) Casimir interaction. 
Note also that here we focus on the average value of the fluctuation-induced force, and we do not address the issue of its fluctuations, which was discussed in a similar context in Refs.~\cite{Bartolo_2002, Gross_2021_fluct_Cas, Napiorkowski_2022}.

\begin{figure}
    \centering
    \begin{tabular}{cc}
        \includegraphics[width=0.46\linewidth]{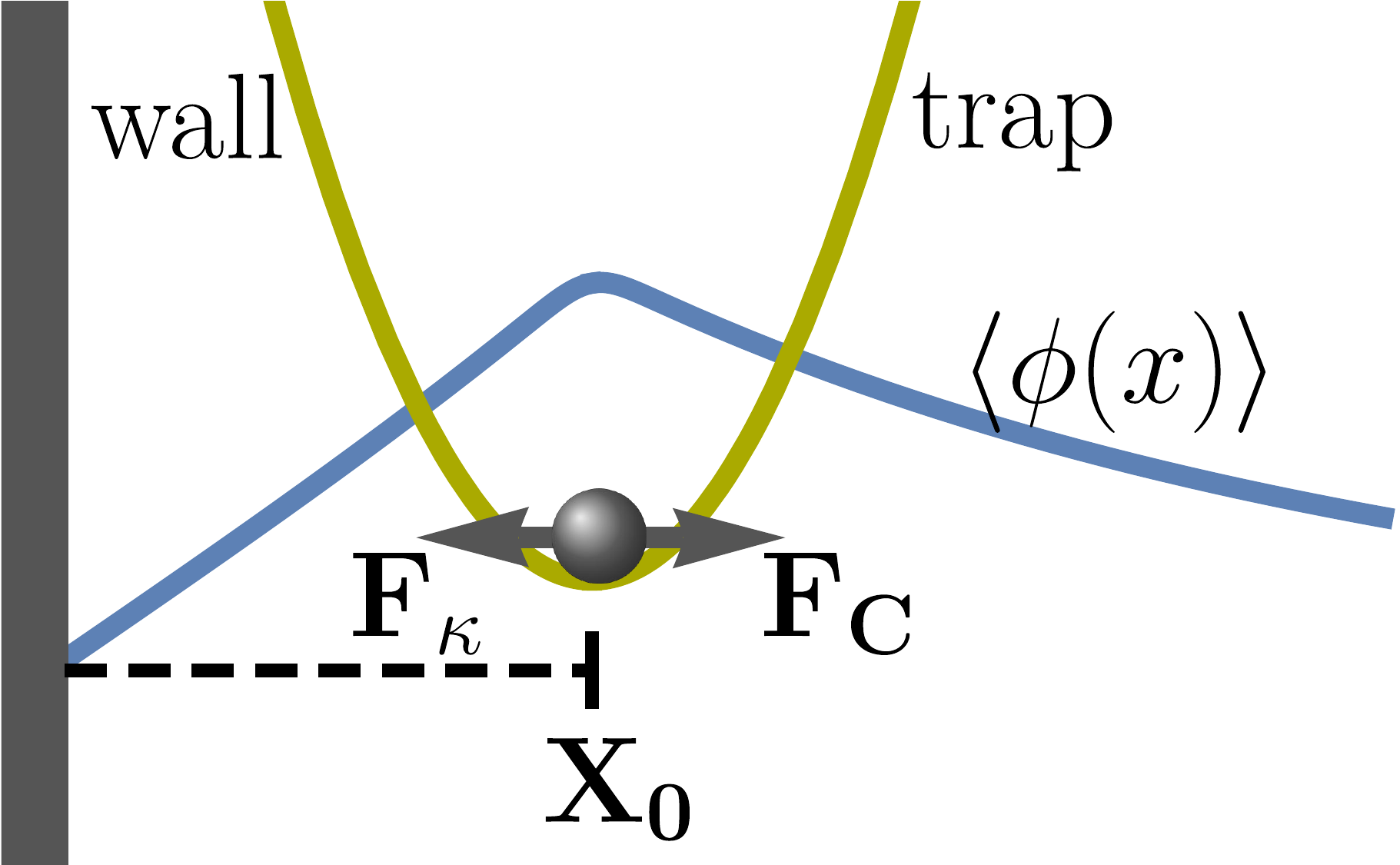} &
        \includegraphics[width=0.49\linewidth]
        {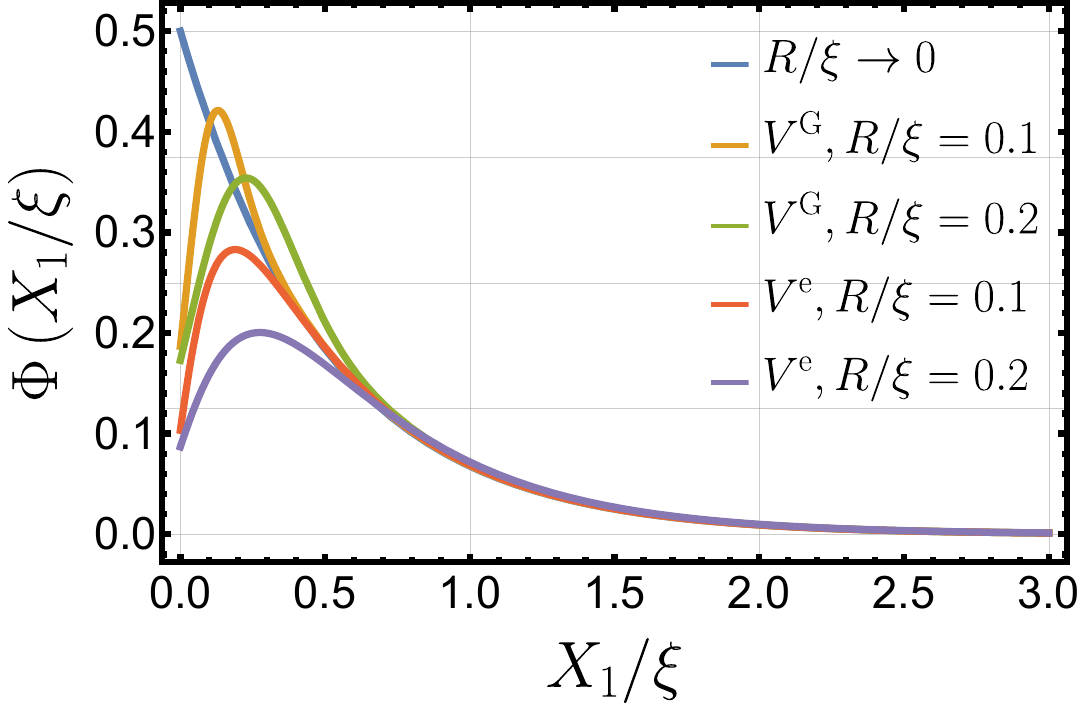} 
         \\
         (a) & (b)
    \end{tabular}
\caption{
(a) Cartoon of the model under investigation: it comprises the tracer particle (black sphere), spatially confined by an optical trap (providing the potential represented by the dark yellow line) and interacting with a fluctuating correlated field $\phi$ (with the dark blue line respesenting its average) which vanishes at the wall according to the boundary condition \eqref{BCs}. Due to the coupling \eqref{eq:Hint} between the particle and the field,  assuming $\lambda$ positive (negative), it is energetically favorable for the particle to occupy the position corresponding to the maximum (minimum) of the field.  Due to the Dirichlet boundary conditions \eqref{BCs} at the wall  and the particle acting as a source of the field in \cref{eq:Hint}, this position is farther away from the wall than the minimum of the trap at $X_0$. 
This field gives rise to a repulsive fluctuation-induced force $\mathbf{F_C} = \mathbf{\hat{e}_1}F_{\rm C}$ causing a shift of the particle position, counterbalanced by the force $\mathbf{F}_{\kappa}$ due to the trap.
(b) Plot of the scaling function $\Phi$ introduced in \cref{eq:Cas_scaling}, in the case of $d=1$, for the Gaussian \eqref{eq:Gauss_ker}, the exponential \eqref{eq:exp_ker} and the point-like \eqref{eq:kernel_point_like_lim} kernels modeling the shape of the particle, see \cref{eq:Hint}. Note that, assuming $\xi \gg R$, the different curves coincide for $X_1/\xi \gtrsim 0.5$.  
}
\label{fig:casimir}
\end{figure}

The force in \cref{Cas_def} turns out to be repulsive, i.e., the particle is pushed away from the wall. This is somehow expected: in fact, when boundary conditions are imposed on the field at the surfaces of both bodies, the force is generically attractive when the boundary conditions on the two surfaces are equal but repulsive otherwise \cite{gambassi_critical_2024}. 
Here, the boundary condition at the wall in Eq.~\eqref{BCs} is even under $\phi \mapsto  - \phi$, whereas the coupling to the particle in Eq.~\eqref{eq:Hint} is odd under such a  transformation. Hence, these different symmetries of the field-wall and field-particle interaction hint towards the  effective force being repulsive. 
Figure \ref{fig:casimir}(a) presents a cartoon of the physical system in which we indicate the forces acting on the colloid. 
We note that, from Eqs.~\eqref{V_eff}, \eqref{propagators}, and \eqref{Cas_def}, one can show that $F_{\rm C}$ satisfies a scaling relation of the form
    \begin{equation}
       F_{\mathrm{C}}(X_1) = \frac{\lambda^2}{X_1^{d-1}} \Phi\left(\frac{X_1}{\xi}, \frac{R}{\xi}\right),
       \label{eq:Cas_scaling}
    \end{equation}
where  $\Phi$ is a dimensionless scaling function. In fact, the prefactor on the r.h.s.~of this equation has the dimension of a force, i.e., $[\lambda^2/X_1^{d-1}] = \mathcal{E}/\mathcal{L}$, which follows from the discussion under \cref{eq:kernel_point_like_lim}. 
At the bulk critical point, $\xi \to \infty$ and the Casimir force exhibits an algebraic behavior as a function of the distance $X_1$ from the wall, with $F_{\mathrm{C}}(X_1) \sim X_1^{-(d-1)}$.
When the field interacts with both bodies via boundary conditions at their surfaces, 
the critical Casimir force acting on parallel planar walls separated by a distance $L$ scales as $F_{\rm C} \sim L^{-d}$ at the critical point, see, e.g., Ref.~\cite{Fisher_1978}. If, instead of two walls, one considers a single wall and a sphere of radius $R$ at a distance $X\gg R$ from it, the resulting fluctuation-induced force can be evaluated by means of the so-called small-sphere expansion  \cite{Burkhardt_1995, Eisenriegler_1995,  Hanke_1998,  Schlesener_2003}. 
The resulting dependence on the distance $X$ turns out to depend on the set of boundary conditions considered. In particular, when the boundary condition at at least one of the two surfaces is invariant under the transformation $\phi \mapsto -\phi$, the force scales as \cite{Burkhardt_1995}
    \begin{equation}
        F_{\rm C}(X) \sim \frac{1}{X^{d + 1 - 1/\nu}},
        \label{eq:Cas_scale_small_sphere}
    \end{equation}
where $\nu$ is the critical exponent introduced above \cref{def:Uk_rec}. We note that in the case of the Gaussian field theory, $\nu = 1/2$ and the scaling in \cref{eq:Cas_scale_small_sphere} yields $F_{\rm C}(X) \sim X^{-(d-1)}$, which is actually the same as in \cref{eq:Cas_scaling}.
Recall that in the current study, the boundary condition \eqref{BCs} is invariant under the considered transformation, while the coupling \eqref{eq:Hint} changes sign. Moreover, the assumptions in Eqs.~\eqref{eq:X0_gg_lT} and~\eqref{eq:X0_gg_R} and the repulsive nature of the Casimir force render $X_1/R\gg 1$. Accordingly, the case considered in this work is analogous to that described by the small-sphere expansion with at least one boundary condition invariant under a global change of the sign of the field.
In the limit $R\to 0$ of a point-like particle, the scaling functions $\Phi$ of the Casimir force take particularly simple forms
 \begin{equation}
     \Phi\left(\frac{X_1}{\xi}, 0 \right) = \begin{cases}
         \frac{1}{2}e^{-\frac{2X_1}{\xi}} \xrightarrow[\xi \to \infty]{} \frac{1}{2},&\mbox{for}  \quad  d=1,\\ 
         \frac{1}{4 \pi} \frac{2X_1}
         {\xi}K_1\left(\frac{2 X_1}{\xi}\right) \xrightarrow[\xi \to \infty]{} \frac{1}{4 \pi }, &\mbox{for}  \quad  d=2,
         \\ \frac{1}{16 \pi}e^{-\frac{2X_1}{\xi}}\left(1 + \frac{2X_1}{\xi} \right) \xrightarrow[\xi \to \infty]{} \frac{1}{16 \pi},&\mbox{for}  \quad d=3,
     \end{cases}
     \label{eq:Cas_f_pointlike}
 \end{equation} 
where we also indicated the expressions at criticality.
Note that in \cref{eq:Cas_f_pointlike}, the exponential decay of the force upon increasing $X_1$ is governed by $2X_1/\xi$. Such a dependence on \textit{twice} the distance from the wall was also observed in previous studies on critical Casimir forces within the small sphere expansion approach \cite{Eisenriegler_1995, Burkhardt_1995}. 
Here, this fact can be understood by interpreting
the boundary condition in Eq.~\eqref{BCs} as being imposed by the presence of a \textit{mirror image particle} located at the coordinate $-X_1$ along the direction orthogonal to the plane, with the coupling constant $-\lambda$. The distance between the original particle and this mirror image is then $2 X_1$.

In \cref{fig:casimir}(b), we plot the scaling function $\Phi$ defined in \cref{eq:Cas_scaling} for $d=1$ and for the various forms of the interaction kernel in Eq.~\eqref{eq:Hint}, i.e., for the Gaussian \eqref{eq:Gauss_ker}, the exponential \eqref{eq:exp_ker}, and for the point-like \eqref{eq:kernel_point_like_lim} kernel, with the corresponding expressions for $R \to 0$ reported in \cref{eq:Cas_f_pointlike}. 
We note that the different curves
actually collapse onto each other for $X_1 \gtrsim \xi/2$ (recall the assumptions in Eqs.~\eqref{eq:X0_gg_lT} and~\eqref{eq:X0_gg_R}). 
This fact is a manifestation of the universality emerging in the system upon approaching its bulk critical point.
In view of this emerging universality, the behavior of the tracer particle trapped sufficiently far from the wall can be conveniently studied by focusing on the case of the point-like potential in Eq.~\eqref{eq:kernel_point_like_lim}, as we do in what follows.

The fluctuation-induced force described above is repulsive in the sense that it pushes the particle away from the wall.
Naturally, this influences the average position  $\langle \mathbf{X} \rangle$ of the particle, causing its shift such that $\mathbf{F}_\mathrm{C}$ is counterbalanced by the resulting force ${\mathbf F}_\kappa$ exerted by the trap, see Fig.~\ref{fig:casimir}(a). As it turns out, the Casimir interaction also affects its variance $\langle \mathbf{X}^2 \rangle - \langle \mathbf{X} \rangle^2$. Noting that $V_{\mathrm{eff}}(\mathbf{x})$ is $\mathcal{O}(\lambda^2)$, one can perform a perturbative calculation in the field-particle coupling $\lambda$ for the moments of the probability distribution of the particle position $X_1$. The derivation is presented in \ref{app_pert_eq}, while we report here the final expression in $d=1$: 
		\begin{align}
			\langle X \rangle  &= X_0 + \frac{\lambda^2}{2 \kappa} e^{-\frac{2 X_0}{\xi} + \frac{2\ell_T^2}{\xi^2}} + \mathcal{O}(\lambda^4),
   \label{1p_eq_1d} \\ 
      \langle X^2 \rangle - \langle X \rangle^2 &= \ell_T^2 \left( 1 -\frac{ \lambda^2}{\kappa \xi}e^{-\frac{2 X_0}{\xi} + \frac{2\ell_T^2}{\xi^2}} \right) + \mathcal{O}(\lambda^4), 
      \label{2p_eq_1d}
 \end{align}
where $\ell_T$ is the thermal length defined in \cref{eq:lt_def}. We note that the factor $ \lambda^2/\kappa$ has the dimension of a distance, see the discussion after \cref{eq:kernel_point_like_lim}.
The presence of the Casimir force reduces the amplitude of the thermal fluctuations of the position of the colloid. In fact, $\mathbf{F}_\mathrm{C}$ pushes the particle further away from the wall against the harmonic trap and it effectively strengthens its confinement, renormalizing the  trap strength $\kappa$ as
\begin{equation}
    \kappa' = \kappa\left(1 + \frac{ \lambda^2}{\kappa\xi}e^{-\frac{2 X_0}{\xi} + \frac{2 \ell_T^2}{ \xi^2}} \right) + \mathcal{O}(\lambda^4).
    \label{kappa_renorm}
\end{equation}
Accordingly, up to terms of order $\lambda^2$,  the variance in Eq.~\eqref{2p_eq_1d} is characterized by the renormalized thermal length $\ell
 '_T= \sqrt{T/\kappa'}<\ell_T$. 
Note that at the critical point ($\xi \to \infty$), the force given by Eqs.~\eqref{eq:Cas_scaling} and \eqref{eq:Cas_f_pointlike} is constant, the average shift in the particle position~\eqref{1p_eq_1d} is maximal, equal to $\lambda^2/(2 \kappa)$, while the fluctuations of the particle position (see \cref{2p_eq_1d}) are unaffected by this additional Casimir-like interaction.  

We conclude by noting that a similar shift in the average position of the particle and an effective tightening of the confinement in the direction perpendicular to the wall is expected in spatial dimensions $d>1$.

\section{Effective nonlinear dynamics of the particle}
\label{sec_eff_dyn}

\subsection{Effective dynamics}
The dynamics of the Gaussian field in Eq.~\eqref{dyn_both1} is linear and therefore it can be integrated. The resulting expression for $\phi(\mathbf{x},t)$ can be substituted into the equation of motion \eqref{dyn_both2} of the particle, which yields the following effective non-Markovian dynamics of the tracer \cite{Demery_2011, Demery_2014, Demery_2019, Basu_2022, Venturelli_2022}:
\begin{equation}
			\gamma_0  \dot{\mathbf{X}}(t) = -\kappa[\mathbf{X}(t) - \mathbf{X}_0] + \int_{-\infty}^t \!\!\mathrm{d} t' \, \mathbf{F}\left(\mathbf{X}(t), \mathbf{X}(t'), t - t'\right) + \mathbf{\Xi}\left(\mathbf{X}(t), t\right) + \bm{\eta}(t),  
            \label{eff_dyn1}
		\end{equation}
where the components of the emerging non-linear 
friction $\mathbf{F}\left(\mathbf{x}, \mathbf{y}, t\right)$ are given by \begin{align}    
			F_i\left(\mathbf{x}, \mathbf{y}, t\right) =-\lambda^2     \int_{\Lambda} \mathrm{d}^d\mathbf{x'}  \int_{\Lambda} \mathrm{d}^d\mathbf{y'} \left[\partial_{x'_i}V(\mathbf{x'} - \mathbf{x})\right]R_{\phi}^D\left(\mathbf{x'}, \mathbf{y}', t \right) V(\mathbf{y'} - \mathbf{y}).
            \label{f_funct} 
\end{align}
In this expression, $R_{\phi}^D(\mathbf{x'}, \mathbf{y'}, t)$ is the response function  of the Gaussian field subject to the Dirichlet boundary condition at the wall, which vanishes
whenever $\mathbf{x}'$ or $\mathbf{y}'$ belong to the boundary. In terms of the bulk response function 
\begin{align}
    R_{\phi}\left(\mathbf{x}- \mathbf{y}, t \right) =  \int \frac{\mathrm{d}^d\mathbf{k}}{(2 \pi)^d} D e^{-D(r+k^2)t-i \mathbf{k} \cdot (\mathbf{x}-\mathbf{y})},
    \label{response2}
\end{align}
$R_{\phi}^D(\mathbf{x'}, \mathbf{y'}, t)$ in Eq.~\eqref{f_funct} is expressed as
\begin{align}
    R_{\phi}^D\left(\mathbf{x}, \mathbf{y}, t \right) = R_{\phi}\left(\mathbf{x}- \mathbf{y}, t \right) -R_{\phi}\left(\mathbf{x}- \mathbf{y}_R, t \right),
    \label{response1}
\end{align}
where the reflected vector $\mathbf{y}_R$ is defined as explained after \cref{Dpropagators}.

Due to the presence of the wall, $\mathbf{F}\left(\mathbf{x}, \mathbf{y}, t\right)$ is not invariant under spatial translation, i.e., it is not a function of 
$\mathbf{x}-\mathbf{y}$, in contrast with the case of the system in the absence of the wall, see Ref.~\cite{Basu_2022}.
Additionally, the field-particle interaction induces a noise term $\mathbf{\Xi}(\mathbf{x},t)$ with vanishing average and variance
            \begin{equation}
			\left\langle \Xi_i(\mathbf{x}, t) \Xi_j(\mathbf{y}, t' ) \right\rangle = T\, G_{ij}(\mathbf{x}, \mathbf{y}, |t-t'|),
            \label{eq:eff_dyn_noise}
		\end{equation}
        where
        \begin{align}    
        G_{ij}(\mathbf{x}, \mathbf{y}, t) =\lambda^2    \int_{\Lambda} \mathrm{d}^d\mathbf{x'}  \int_{\Lambda} \mathrm{d}^d\mathbf{y'} \left[\partial_{x'_i}V(\mathbf{x'} - \mathbf{x})\right] C_{\phi}^D\left(\mathbf{x'}, \mathbf{y}', t \right)  \left[\partial_{y'_j}V(\mathbf{y'} - \mathbf{y})\right]. \label{g_funct}
        \end{align}
Here, $C^D_\phi$ is the correlator of the Gaussian field satisfying Dirichlet boundary conditions 
\begin{align}
&C_{\phi}^D\left(\mathbf{x}, \mathbf{y}, t \right) = C_{\phi}\left(\mathbf{x} - \mathbf{y}, t \right) - C_{\phi}\left(\mathbf{x} - \mathbf{y}_R, t \right),
\label{C_phi1}
\end{align}
where $\mathbf{y}_R$ is the reflected vector defined after \cref{Dpropagators}, while
\begin{align}    
C_{\phi}\left(\mathbf{x}, t \right) =  \int \frac{\mathrm{d}^d\mathbf{k}}{(2 \pi)^d} \frac{1}{r + k^2} e^{-D(r+k^2)t-i \mathbf{k} \cdot \mathbf{x}}, 
\label{C_phi2}
\end{align}
is the correlator of the Gaussian field in the bulk, 
with $C_{\phi}\left(\mathbf{x}, t = 0 \right) = C_{\phi,\textrm{st}}(\mathbf{x})$ and thus $C^D_{\phi}\left(\mathbf{x}, \mathbf{y}, t = 0 \right) = C^D_{\phi,\textrm{st}}(\mathbf{x}, \mathbf{y})$, see Eqs.~\eqref{Dpropagators} and \eqref{propagators}.
Moreover, we note that 	
\begin{equation}
			\int_{-\infty}^t \!\!\mathrm{d} t'\,\mathbf{F}(\mathbf{x}, \mathbf{x}, t - t') = \mathbf{\hat{e}}_1\,  F_{\mathrm{C}}\left(x_1\right) \equiv \mathbf{F}_{\mathrm C},
            \label{Cas_from_F}
		\end{equation}
where $\mathbf{F}_{\mathrm{C}}$ is the Casimir force in \cref{Cas_def}, acting on the particle and directed along $\mathbf{\hat{e}}_1$ normal to the surface of the wall. 
This equality can be understood as follows: if the particle is pinned at a certain position $\mathbf{x}$, then the average force that the fluctuating field exerts on it at that position can be read from the r.h.s.~of Eq.~\eqref{eff_dyn1} with $\mathbf{X}(t)=\mathbf{x}$ and it is therefore given by 
$\int_{-\infty}^t \!\!\mathrm{d} t'\,\mathbf{F}(\mathbf{x}, \mathbf{x}, t - t')$, i.e., by the l.h.s.~of Eq.~\eqref{Cas_from_F}. 
On the other hand, the average force exerted on a fixed particle by the field is nothing but the static fluctuation-induced force $F_{\mathrm{C}} (x_1)$, from which Eq.~\eqref{Cas_from_F} follows immediately. 
Accordingly, by expressing in Eq.~\eqref{eff_dyn1} the friction $\mathbf{F}(\mathbf{x}, \mathbf{y}, t)$ as the sum of its ``off-diagonal'' component 
\begin{equation}
			\mathbf{\hat{F}}(\mathbf{x}, \mathbf{y}, t) \equiv \mathbf{F}(\mathbf{x}, \mathbf{y}, t) - \mathbf{F}(\mathbf{x}, \mathbf{x}, t)
        \label{decomp}
		\end{equation}
and its diagonal component $\mathbf{F}(\mathbf{x}, \mathbf{x}, t)$ allows us to isolate the field-induced non-linear non-Markovian friction $\hat{\mathbf{F}}$ in the tracer dynamics. 
The resulting equation of motion of the particle then contains a term which is explicitly due to the Casimir force at equilibrium: 
    \begin{align}
			\gamma_0 \dot{\mathbf{X}}(t) = -\kappa\left[\mathbf{X}(t) - \mathbf{X}_0\right]
            +  \mathbf{F}_{\mathrm{C}}\left(X_1(t)\right) 
            + \int_{-\infty}^t \!\!\mathrm{d} t' \, & \mathbf{\hat{F}}\left(\mathbf{X}(t), \mathbf{X}(t'), t - t'\right)   \n \\ &+ \mathbf{\Xi}\left(\mathbf{X}(t), t\right) + \bm{\eta}(t). 
   \label{eff_dyn2}
		\end{align}
We note that the effective dynamics in Eq.~\eqref{eff_dyn1} (or Eq.~\eqref{eff_dyn2}) is exact in the sense that it describes the position of the particle for any value of the coupling constant $\lambda$ of the field-particle interaction in Eq.~\eqref{eq:Hint}.
Notably, upon increasing the timescale separation between the dynamics of the particle and of the field --- which is obtained in the \emph{adiabatic limit} $D \to \infty$ with $r$ finite, --- $\bm{\Xi}$ and   $\bm{\hat{F}}$ vanish, yielding a Markovian evolution of the position of the particle. 
Then, the particle-field interaction manifests itself only in the presence of the Casimir-like force, and the effective dynamics takes the form
 		\begin{equation}
			\gamma_0 \dot{\mathbf{X}}(t) = -\kappa\left[\mathbf{X}(t) - \mathbf{X}_0\right] + \mathbf{F}_{\mathrm{C}}\left(X_1\right) + \bm{\eta}(t),
            \label{eff_dyn_markov}
		\end{equation}
which is the  Langevin equation for a particle immersed in a simple fluid interacting via fluctuation-induced forces with the wall.
We refer the reader to~\ref{ad_app} for more details concerning the adiabatic limit.

\subsection{Fluctuation-dissipation relation}
        
The function  $\mathbf{F}$ given by Eq.~\eqref{f_funct}, describing the field-induced friction and the Casimir-like force acting on the particle, and the matrix $\mathbb{G}$ of components $G_{ij}$ given by Eq.~\eqref{g_funct} and describing the correlations of the field-induced noise are related to each other via
         \begin{equation}
                  - \frac{\partial}{\partial t} G_{ij}(\mathbf{x}, \mathbf{y}, t) = \frac{\partial}{\partial y_j} F_i(\mathbf{x}, \mathbf{y}, t)
                  \quad \mbox{for}\quad t>0.
                  \label{FDT_eff}
            \end{equation}
As we show in~\ref{app_t_reversal}, whenever this relation holds, the dynamics is invariant under time-reversal. 
This means that the stationary state of this process is, in fact, the equilibrium one. Accordingly,  the property~(\ref{FDT_eff}) is a manifestation of the fluctuation-dissipation relation for the nonlinear, non-translationally invariant dynamics considered in~\cref{eff_dyn1}. The result above is a generalization of such a  relation for translationally-invariant systems,  derived in Ref.~\cite{Basu_2022}.

\subsection{The linearized memory kernel}

In order to relate the nonlinear friction $\mathbf{\hat{F}} (\mathbf{x}, \mathbf{y}, t)$  with the usual linear memory kernel, let us denote $\mathbf{X}(t) - \mathbf{X}(t') =  \mathbf{\Delta}(t,t')$ and assume that $|\mathbf{\Delta}(t,t')|$ is sufficiently small to justify treating it as an expansion parameter.  We use the fluctuation-dissipation relation in Eq.~\eqref{FDT_eff} to write
                \begin{align}    
                    &\hat{F}_i(\mathbf{X}(t), \mathbf{X}(t'), t-t')  \simeq -\Delta_j(t, t') \partial_{y_j}F_i(\mathbf{x}, \mathbf{y}, t-t')\vert_{\mathbf{x} = \mathbf{y} = \mathbf{X}(t)}  \n   \\ &= \Delta_{j}(t, t') \partial_tG_{ij}(\mathbf{x}, \mathbf{y}, t-t')\vert_{\mathbf{x} = \mathbf{y} = \mathbf{X}(t)} =  -\Delta_{j}(t, t') \frac{\mathrm{d}}{\mathrm{d} t'}G_{ij}(\mathbf{X}(t), \mathbf{X}(t), t-t'),
                \end{align}
where we employed the Einstein summation convention. This relation allows us to rewrite the friction term in Eq.~\eqref{eff_dyn2} in the form
            \begin{align}  
            \int_{-\infty}^t \!\!\mathrm{d}t' \, \hat{F}_i(\mathbf{X}(t), \mathbf{X}(t'), t-t')  & \approx - \int_{-\infty}^t \!\!\mathrm{d}t' \frac{\mathrm{d}}{\mathrm{d} t'}\left[G_{ij}(\mathbf{X}(t), \mathbf{X}(t), t-t')\right]\Delta_j(t,t')  \n \\  &= \int_{-\infty}^t \!\!\mathrm{d}t'\, G_{ij}(\mathbf{X}(t), \mathbf{X}(t), t-t')\frac{\mathrm{d}}{\mathrm{d} t'}\left[\Delta_j(t,t')\right]  \n \\ &=  - \int_{-\infty}^t \!\!\mathrm{d}t'\, G_{ij}(\mathbf{X}(t), \mathbf{X}(t), t-t'){\dot X}_j(t'),
            \end{align}    
where the last line has the usual form of a non-Markovian (linear) friction.
Moreover, it follows from Eq.~\eqref{g_funct} that $\mathbb{G}(\mathbf{x}, \mathbf{x}, t)$ depends only on the first component $x_1$ of $\mathbf{x}$.  
Accordingly, the linearization of $\mathbf{F}(\mathbf{x}, \mathbf{y},t)$ in $\mathbf{x} - \mathbf{y}$ results into a linear memory kernel $\mathbf{\Gamma}$ (see \cref{sec:intro}) with components $\Gamma_{ij}$ which depends on the distance from the wall, given by
            \begin{equation}
                \Gamma_{ij}(x_1, t) = G_{ij}(\mathbf{x}, \mathbf{x}, t). 
            \end{equation}
Also, at the lowest order in $|\mathbf{x} - \mathbf{y}|$, 
            \begin{equation}
            \langle \Xi_i(\mathbf{x}, t) \Xi_j(\mathbf{y}, t') \rangle = T \,  G_{ij}(\mathbf{x}, \mathbf{y},  |t - t'|) \approx  T \,  G_{ij}(\mathbf{x},\mathbf{x},  |t - t'|) = \Gamma_{ij}(x_1, t),
            \end{equation}  
which is the usual fluctuation-dissipation relation for the linear stochastic dynamics \cite{Kubo_book}.
Note that both the memory kernel and, consequently, the time correlation of the noise are state dependent since they depend on $X_1$. Accordingly, the linearized version of the dynamics \eqref{eff_dyn1} features a multiplicative field-induced noise. This is in contrast with the bulk case \cite{Basu_2022}, in which the linearization of the corresponding dynamics yields an additive noise term. 

\section{Perturbative calculation of the correlation functions}
\label{sec_pert_exp}

\subsection{Perturbative expansion}

Solving the system of equations (\ref{dyn_both1}) and (\ref{dyn_both2}) allows us to describe the dynamical behavior of the colloidal particle. However, due to the nonlinearity in $\{\phi, \mathbf{X}\}$ of the dynamics stemming from the field-particle coupling $\propto \lambda$, it is not solvable in general. Thus, we resort to a perturbative expansion in the coupling constant
		\begin{align}
				&\phi(\mathbf{x}, t) = \phi^{(0)}(\mathbf{x}, t) + \lambda \phi^{(1)}(\mathbf{x}, t) + \lambda^2 \phi^{(2)}(\mathbf{x}, t) + \mathcal{O}(\lambda^3), 
			\label{pert_exp_def_field}	\\
				&\mathbf{X}(t) = \mathbf{X}^{(0)}(t) + \lambda \mathbf{X}^{(1)}(t) + \lambda^2 \mathbf{X}^{(2)}(t)  + \mathcal{O}(\lambda^3).
                \label{pert_exp_def_part}
		\end{align}
Here, $\phi^{(0)}(\mathbf{x}, t)$ and $\mathbf{X}^{(0)}(t)$ are the solutions of Eqs.~(\ref{dyn_both1}) and (\ref{dyn_both2}) when the tracer particle and the field are decoupled, i.e., for $\lambda = 0$. In Refs.~\cite{Demery_2011, Demery_2014, Demery_2019}, 
a similar perturbative analysis was performed within the path-integral framework for the generating function of the dynamics. Here, as in Refs.~\cite{Basu_2022, Demery_Gabassi_2023, Venturelli_2022, Venturelli_2022_2parts, Venturelli_2023, Venturelli_2024, Pruszczyk_2025}, we carry out a direct calculation based on the perturbative expansion of the dynamical equations. 
As indicated in Eqs.~(\ref{pert_exp_def_field}) and (\ref{pert_exp_def_part}), we will concern ourselves with the expansion up to the quadratic order in $\lambda$, as it will prove itself sufficient to capture the non-trivial effects on the dynamics of the tracer particle stemming from the coupling to the slow-relaxing field. 
We focus on the stationary properties of the system, i.e.,  we consider the case in which the initial conditions of the dynamics were specified at time $t_0 \to -\infty$, which renders them irrelevant. Moreover, unless stated otherwise, we shall focus below on the one-dimensional problem $d=1$.
   
Details of the derivations of the predictions presented below are reported in \ref{app_pert_exp}.
In particular, in~\ref{1point_app}, we show that perturbatively solving the equations of motion \eqref{dyn_both1} and \eqref{dyn_both2} renders the same steady-state average position of the particle  reported in Eq.~\eqref{1p_eq_1d} obtained from the Gibbs-Boltzmann distribution.

\subsection{Two-time correlation function}
\label{subsec:corr}

We focus here on the perturbative calculation of the connected two-time correlation function $C_c(t)$. Studying 
$C_c(t)$ will not only reveal the effects of the Casimir-like force on the time correlations of the process, 
but also how  the presence of the wall affects the nonlinear memory previously derived  in Ref.~\cite{Basu_2022} for a particle in the bulk (i.e., for $X_0 \to \infty$). 

\begin{figure}
    \centering
    \begin{tabular}{cc}
        \includegraphics[width=0.47\linewidth]{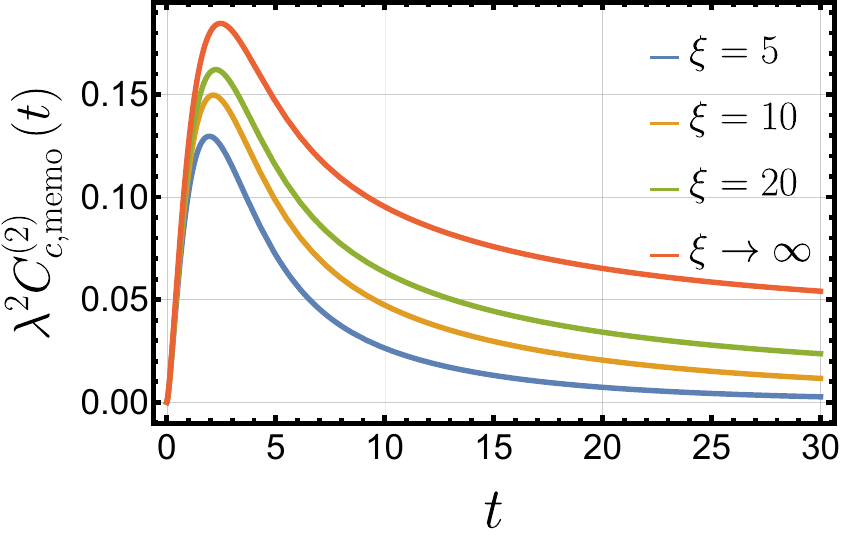}&
        \includegraphics[width=0.47\linewidth]{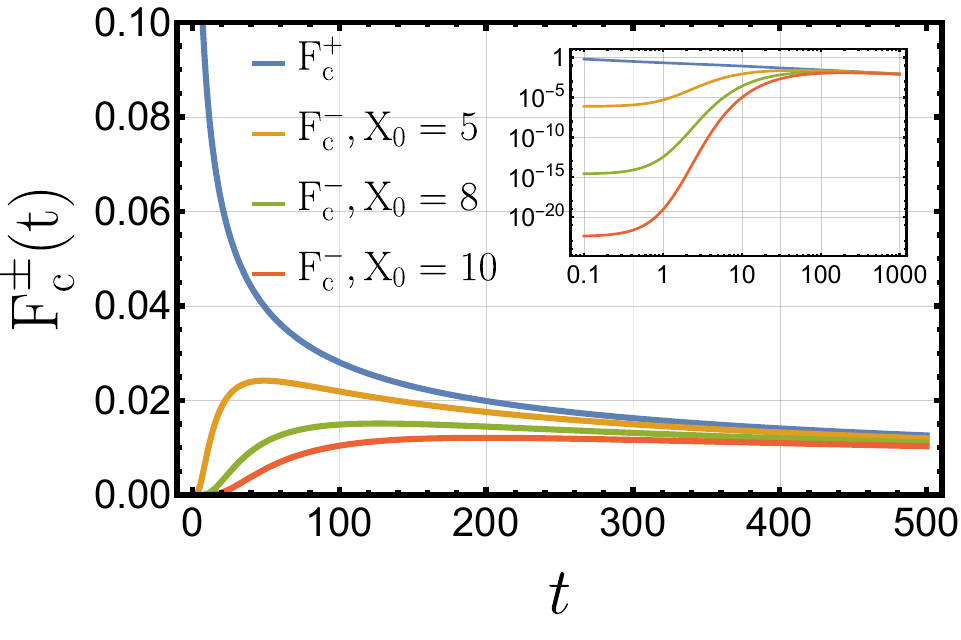}
        \\[-2mm]
         (a) & (b) 
    \end{tabular}
\caption{(a) Memory-induced contribution $\lambda^2 C_{c, \mathrm{memo}}^{(2)}(t)$ to the correlation  function in Eq.~\eqref{C_c_mem} as a function of time $t$ for $X_0=10$.  Note that this contribution vanishes at $t=0$ since it takes a finite amount of time for the memory to ``build up''. After an initial growth $\sim t^{3/2}$ stemming from the $\sim t^{-1/2}$ behavior of the memory \eqref{mem_small_t} integrated in \cref{C_c_mem},
$\lambda^2 C_{c, \mathrm{memo}}^{(2)}(t)$ attains its maximal value at $t \approx 2$ and then, upon further increasing $t$,  it  decreases either  exponentially for finite $\xi$, or algebraically $\sim t^{-1/2}$ for $\xi \to \infty$. 
(b) Bulk ($\mathrm{F^+}$) and wall ($\mathrm{F^-}$)  
critical contributions to the 
memory given by Eqs.~\eqref{memo_crit_pl} and \eqref{memo_crit_min}, respectively, as functions of time $t$. We note that $\mathrm{F^+}$ and $\mathrm{F^-}$ coincide at long times.  In the inset, we plot $\mathrm{F^\pm}$ on a log-log scale in order emphasize the different order of magnitudes of the bulk and the wall contributions at short times. 
Quantities in this figure are plotted for $\lambda^2/(2 \kappa) = \ell_T = \sqrt{D/\omega_0} =1$. 
}
\label{fig:noncrit_memo}
\end{figure}

Since the equations of motion \eqref{dyn_both1} and \eqref{dyn_both2} are invariant under the transformation \eqref{parity}, the two-time correlation function is an even function of $\lambda$, and the linear contribution vanishes:
		\begin{equation}
			C_c(t_1, t_2) =\langle X(t_1) X(t_2) \rangle - \langle X \rangle^2 = C^{(0)}_c(t_1, t_2) + \lambda^2  C^{(2)}_c(t_1, t_2) + \mathcal{O}(\lambda^4).
		\end{equation}
For $\lambda=0$, one recovers the Ornstein-Uhlenbeck process
		\begin{equation}
			C^{(0)}_c(t_1, t_2) = \ell_T^2\, e^{-\omega_0 t}\quad \mbox{with} \quad  t = |t_1 - t_2|,
   \label{C_c_0}
		\end{equation}
where $\ell_T$ is the thermal lengthscale defined in \cref{eq:lt_def}, and  $\omega_0 = \kappa/\gamma_0$ is the relaxation rate of the colloid in the harmonic trap. The details of the calculation of $\lambda^2C^{(2)}_c(t)$ are presented in~\ref{corr_app}, while here we report the final expression  
        \begin{equation}
            \lambda^2 C^{(2)}_c(t) = \lambda^2 C_{c,\mathrm{C}}^{(2)}(t) + \lambda^2 C_{c, \mathrm{memo}}^{(2)}(t),
            \label{C_c_2}
        \end{equation}
        where 
        \begin{equation}
            \lambda^2 C_{c,\mathrm{C}}^{(2)}(t) =  -\frac{\lambda^2\ell_T^2}{\kappa\xi} e^{- \frac{2 X_0}{\xi} + \frac{\ell_T^2}{ \xi^2}}e^{- \omega_0 t}(1 + \omega_0 t)
            \label{C_c_C}
        \end{equation}
        is the Casimir contribution and
		\begin{equation}
			\lambda^2 C_{c, \mathrm{memo}}^{(2)}(t)  =     \frac{\lambda^2\ell_T^2}{ \kappa \xi}  \int_0^{\omega_0 t} \!\mathrm{d} u\, ( \omega_0 t -u)e^{- (\omega_0 t - u)} \mathcal{F}(u) 
            \label{C_c_mem}
            \end{equation}
is the memory-induced  contribution to the two-point function. We note that the presence of both terms above is expected for any $d \geq 1$. In the equation above, $\mathcal{F}$ denotes the emergent dimensionless memory 
\begin{align}
    \mathcal{F}(u) &=  \int \frac{\mathrm{d}v}{2 \pi}\frac{v^2}{1 + v^2}e^{-(D/\omega_0\xi^2)u(1 + v^2)}e^{-(\ell_T/\xi)^2v^2(1 - e^{-u})} \n 
     \\  &+ \int \frac{\mathrm{d}v}{2 \pi}\frac{v^2}{1 + v^2}e^{-(D/\omega_0\xi^2)u(1 + v^2)} e^{-(\ell_T/\xi)^2v^2(1 + e^{-u})} \cos \left(2 (X_0/\xi) v \right),
    \label{dimless_memo}
\end{align}
with the closed-form expressions obtained after the integration reported in \ref{corr_app}, c.f., Eqs.~\eqref{dimfull_memo},~\eqref{F+_dimfull_app},~\eqref{F-_dimfull_app},~\eqref{F+_app},~\eqref{F-_app}, and \eqref{dimless_memo_app}.
The function  $\mathcal{F}(u)$ defined above is proportional to  $\left \langle  G\left(X^{(0)}(t+u), X^{(0)}(t) , u\right) \right\rangle$ with $u>0$, where $G$ is given by \cref{g_funct} (we omit here the indication of the indices of $G$, since we consider the case $d=1$, where they take only one value),
and  $\langle \cdot\rangle$ denotes averaging over the stationary Ornstein-Uhlenbeck process. 
We note that at $t=0$, the memory-induced contribution \eqref{C_c_mem} vanishes, and the Casimir contribution \eqref{C_c_C} reduces to the expression reported in Eq.~\eqref{2p_eq_1d}. 
Moreover, the Casimir contribution is independent of the diffusion coefficient $D$ and, therefore, it is not affected by taking the adiabatic limit (see the discussion above \cref{eff_dyn_markov}), whereas the memory term depends on $D$ and vanishes in this limit (see~\ref{ad_app} and  Eq.~\eqref{dimfull_memo} for details). 
Finally, it follows from Eqs.~\eqref{C_c_0} and \eqref{C_c_C} that  renormalizing the relaxation rate according to $\omega_0  \mapsto \omega_0' ={\kappa'}/{\gamma_0}$ with $\kappa'$ defined in Eq.~\eqref{kappa_renorm}, renders 
  \begin{equation}
      	C^{(0)}_c(t) + \lambda^2 C_{c,\mathrm{C}}^{(2)}(t)  = \ell
 '^2_Te^{-\omega_0' t} + \mathcal{O}(\lambda^4),
        \label{2p_OU_renorm}
  \end{equation}
where $\ell_T'$ was introduced after Eq.~\eqref{kappa_renorm}.
We recognize Eq.~\eqref{2p_OU_renorm} as the two-time correlator of the Ornstein-Uhlenbeck process with the renormalized trap strength.
In Fig.~\ref{fig:noncrit_memo}(a), we present a plot of  $\lambda^2 C_{c, \mathrm{memo}}^{(2)}(t)$ for various values of  $\xi$ in the case of $\lambda^2/(2 \kappa) =  \ell_T =   \sqrt{D/\omega_0}=1$  and $X_0=10$. At criticality ($r=0$), its long-time behavior is characterized by an algebraic decay~$\propto t^{-1/2}$. For finite values of $r$, it decays exponentially at long times. We note that both $C_c^{(0)}(t)$ and $\lambda^2 C_{c,\mathrm{C}}^{(2)}(t)$ decay exponentially upon increasing time, irrespective of the value of the parameter $r$. Accordingly, at the critical point, the memory-induced term is the dominant one at long times.

Interestingly enough, the memory-induced term $\lambda^2 C_{c, \mathrm{memo}}^{(2)}(t)$ depends on $X_0$, and the emergent memory $\mathcal{F}$ in Eq.~\eqref{C_c_mem} can be decomposed into two terms
  \begin{equation}
      \mathcal{F}(u) =  \mathcal{F}^+(u) + \mathcal{F}^-(u),
      \label{dimless_mem_decomp}
  \end{equation}
where $\mathcal{F}^+$ does not depend on the wall-trap separation $X_0$, whereas $\mathcal{F}^-$ does, and it vanishes in the bulk, i.e., in the limit $X_0 \to \infty$.
We refer to them as the bulk and the wall contributions to the dimensionless memory, respectively. 
The expressions of $\mathcal{F}^+$ and $\mathcal{F}^-$ for $r >0$ are provided by the first and the second term in Eq.~\eqref{dimless_memo}, respectively. The integration over ${v}$ therein renders rather lengthy closed-form expressions reported in Eqs.~\eqref{F+_app}, \eqref{F-_app}, and \eqref{dimless_memo_app}. 
Here, we focus on the simpler expressions obtained at the critical point $r=0$:
\begin{align}    
		& \frac{1}{\xi} \mathcal{F}^+(\omega_0 t) \xrightarrow[\xi \to \infty]{} \frac{1}{\sqrt{4 \pi \left[Dt + \ell_T^2(1 - e^{-\omega_0 t}) \right]}} = \mathrm{F}_c^+(t), 
        \label{memo_crit_pl}
        \\ 
      		&\frac{1}{\xi} \mathcal{F}^-(\omega_0 t) \xrightarrow[\xi \to \infty]{} \frac{\exp\left[ - \frac{X_0^2}{Dt + \ell_T^2(1 + e^{-\omega_0 t})} \right]}{\sqrt{4 \pi \left[Dt + \ell_T^2(1 + e^{-\omega_0 t}) \right]}}= \mathrm{F}_c^-(t), 
\label{memo_crit_min}
\end{align}
where $\ell_T$ is the thermal length defined in \cref{eq:lt_def}, and we denote by $\mathrm{F}_c^{\pm}(t)$ the  critical contributions to the 
memory.
(Note that in taking the limit $\xi\to\infty$ in the equations above, we have accounted for the fact that $\lambda^2 C_{c, \mathrm{memo}}^{(2)}(t)$ in Eq.~\eqref{C_c_mem} depends on ${\mathcal F}^\pm$ via ${\mathcal F}/\xi$, owing the prefactor $\propto 1/\xi$ of the integral.) 
The plots of $\mathrm{F}_c^{\pm}(t)$ are presented in Fig.~\ref{fig:noncrit_memo}(b).   In the inset, the quantities are plotted on a log-log scale, which highlights the different magnitudes of the wall and the bulk contributions at short times. 
To understand the physical significance of both contributions, we investigate their asymptotic behavior. In particular, for the bulk contribution, we find
\begin{equation}
\mathrm{F}_c^+(t) \approx \begin{cases}
	 \frac{1}{\sqrt{4 \pi[D + ({T}/{\gamma_0})]t}},    &\mbox{for} \quad \omega_0t\ll 1,
    \\ 
      		   \frac{1}{\sqrt{4 \pi(Dt + \ell_T^2)}},    &\mbox{for} \quad \omega_0t\gg 1,
\end{cases} 
\label{mem_small_t}
\end{equation}
while, for the wall contribution, 
\begin{equation}
\mathrm{F}_c^-(t) \approx
\begin{cases}
	 \frac{1}{2\sqrt{2 \pi \ell_T^2}}\exp\left[ - \frac{X_0^2}{2\ell_T^2} \right],  &\mbox{for} \quad \omega_0t\ll 1,
    \\ 
      		\frac{1}{\sqrt{4 \pi(Dt + \ell_T^2)}}\exp\left[ - \frac{X_0^2}{Dt + \ell_T^2} \right],  &\mbox{for} \quad \omega_0t\gg 1.
         \label{mem_large_t}
\end{cases}    
\end{equation}
Note that, within the short-time regime $t\ll \omega_0^{-1}$, $\mathrm{F}_c^+(t)$ displays an integrable  singularity $\propto t^{-1/2}$ (recall that $\mathrm{F}_c^+(t)$ determines the integrand in \cref{C_c_mem}), giving rise to the initial $t^{3/2}$ growth of the memory-induced contribution to the two-point function, 
%
%
while $\mathrm{F}_c^-(t)$ approaches a constant value which, due to the  assumption \eqref{eq:X0_gg_lT}, is very small. 
Moreover, at times $t \gtrsim t^* = (X_0^2 - \ell_T^2)/D \approx X_0^2/D$, the wall contribution becomes comparable with the bulk contribution (see the argument of the exponent in Eq.~\eqref{mem_large_t}), and they coincide as $t \to \infty$. 
The physical interpretation of these observations is as follows: at short times, the position of the particle is strongly correlated with its initial position (testified by the $t^{-1/2}$ singularity  of the memory as $t\to 0$), and the colloid is not significantly  affected by the presence of the wall due to the assumption in Eq.~\eqref{eq:X0_gg_lT}.  However, 
because of the presence of long-range correlations at criticality, the field disturbed by the random motion of the particle is correlated with the field in the vicinity of the wall. The time required for these correlations to propagate is of the order of magnitude of the typical diffusion time across the distance $X_0$ with diffusivity $D$. A similar effect of ``retardation'' of fluctuation-induced forces was discussed in Refs.~\cite{Furukawa_2013, Venturelli_2022_2parts, Fournier_2021}. 
Finally, at very long times, the wall contribution coincides with the bulk contribution. This is because, as we discussed below \cref{eq:Cas_f_pointlike}, 
the imposition of the boundary condition  \eqref{BCs} at the wall can be interpreted as a consequence of the presence of a mirror-image particle at $- X$ coupled to the field with coupling $-\lambda$. At long times, the correlations due to this second ``virtual'' particle are observed.

\begin{figure}
    \centering
    \begin{tabular}{cc}
        \includegraphics[width=0.47\linewidth]{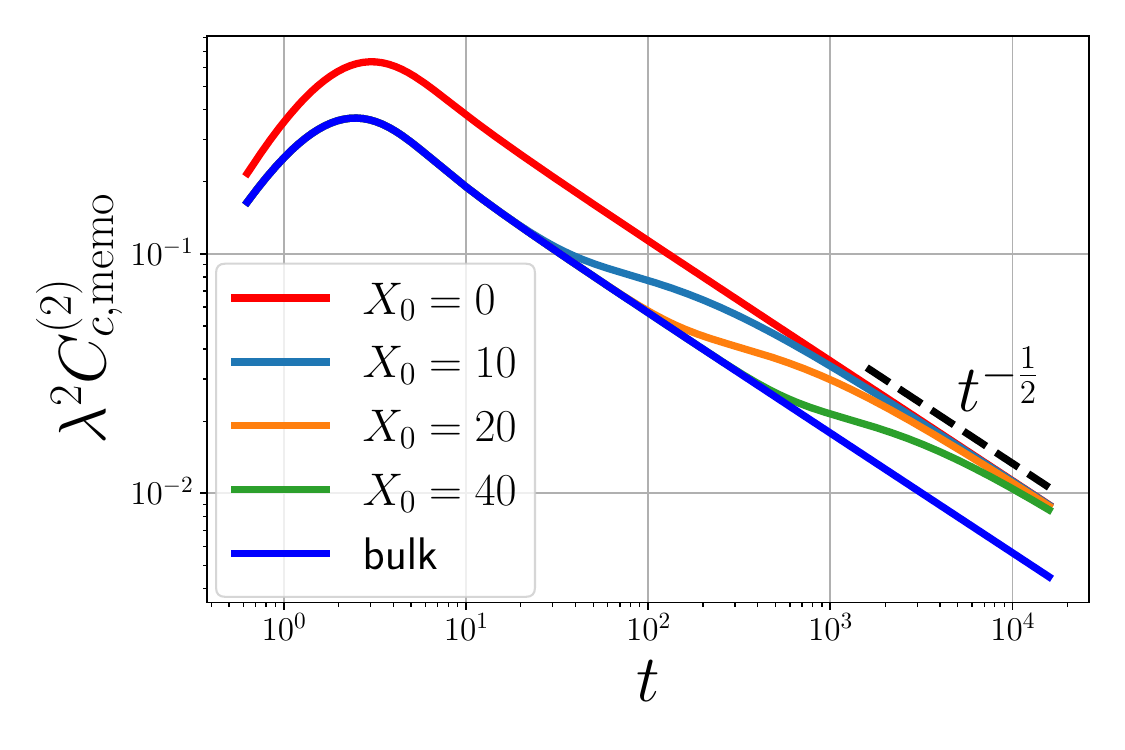} &
        \includegraphics[width=0.47\linewidth]{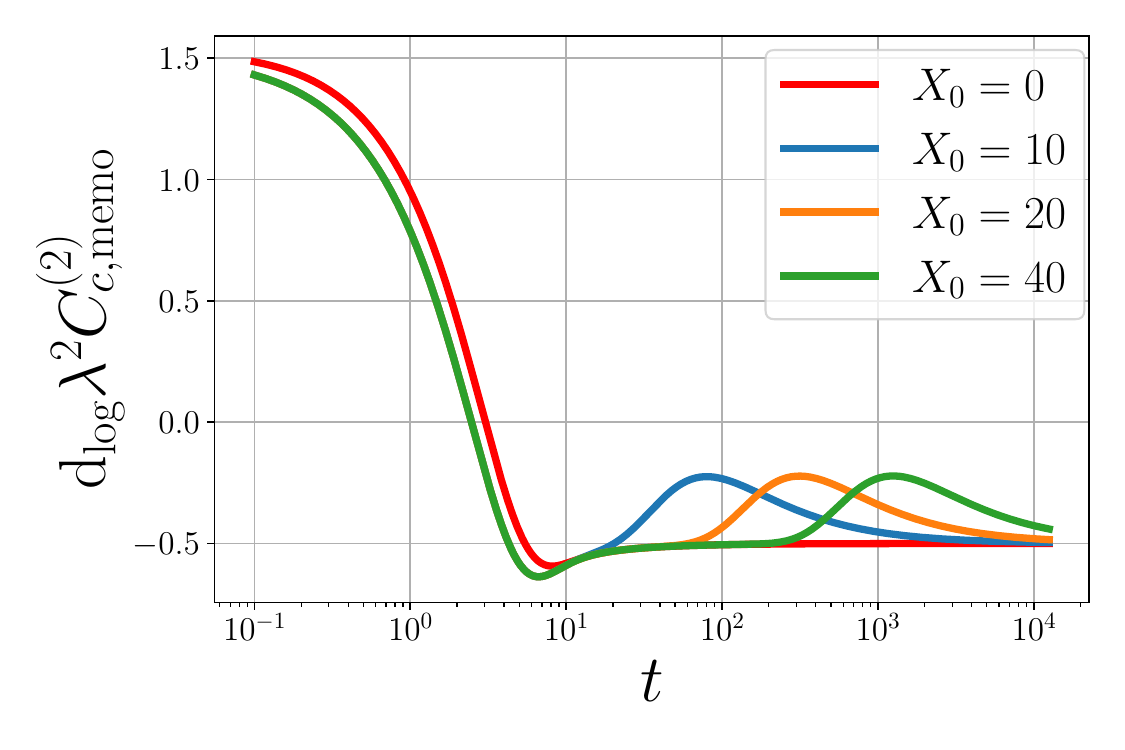}
        \\[-2mm]
         (a) & (b) 
    \end{tabular}
\caption{Memory-induced contribution $\lambda^2 C_{c, \mathrm{memo}}^{(2)}(t)$ to the correlation function in Eq.~\eqref{C_c_mem} at the critical point $r=0$ and as a function of time $t$. 
(a) Plot on a log-log scale of $\lambda^2 C_{c, \mathrm{memo}}^{(2)}(t)$, which highlights the algebraic decay $\propto t^{-1/2}$ at long times. Additionally to the plots obtained for finite values of $X_0$, we present the asymptotic curves corresponding to $X_0 = 0$ (upper curve) and $X_0 \to \infty$ (lower curve). For finite values of $X_0$, one observes a cross over from the bulk asymptotic to the wall asymptotic. 
(b) Logarithmic derivative (see \cref{eq:log_der_def}) of $\lambda^2C_{c, \mathrm{memo}}^{(2)}(t)$ as a function of time $t$. At short times, the values of the logarithmic derivative approaches $3/2$, corresponding to a $\propto t^{3/2}$ growth of the memory-induced contribution, whereas it takes the value $-1/2$ at long times, corresponding to the algebraic behavior discussed above. 
The bumps in the plots indicate the times at which the crossovers from the bulk to the wall asymptotic occur. Note that the larger $X_0$, the longer the crossover time. Both plots correspond to  $\lambda^2/(2 \kappa) = \ell_T = \sqrt{D/\omega_0} =1$. 
}
\label{fig:crit_memo}
\end{figure}

The behavior of the memory discussed above affects the memory-induced contribution to the correlation function in Eq.~\eqref{C_c_mem}. In Fig.~\ref{fig:crit_memo}(a), we plot  $\lambda^2 C_{c, \mathrm{memo}}^{(2)}(t)$ on a log-log scale for $r = 0$ and various values of the distance $X_0$ between the minimum of the trap and the wall. For concreteness, we consider $\lambda^2/(2 \kappa) =  \ell_T =   \sqrt{D/\omega_0}=1$. For comparison, we plot two asymptotic curves: one corresponding to the limit $X_0 \to \infty$, i.e.,  the asymptotic bulk behavior, and one corresponding to $X_0 =0$, which we call the  wall asymptotic. Note that the latter, for $t \gg \omega_0^{-1}$, turns out to be twice as large as the curve corresponding to the case $X_0 \to \infty$, which stems from the correlations with the mirror-image particle at $-{X}$.

For finite distances from the wall, at short times, $\lambda^2 C_{c, \mathrm{memo}}^{(2)}(t)$ follows the bulk asymptotic behavior; at moderate times, it crosses over to the wall asymptotic, which it follows at longer times.  Note that as the wall-particle separation increases, the crossover time also increases, in agreement with  Eq.~\eqref{mem_large_t}.
 To highlight the crossover of $\lambda^2C_{c, \mathrm{memo}}(t)$ from wall asymptotic to bulk asymptotic,  we consider its logarithmic derivative $\mathrm{d}_{\mathrm{log}}$, which, for a (positive) function $f(t)$ is defined as
\begin{equation}
    \mathrm{d}_{\log}f(t) = \frac{\mathrm{d} \log f(t)}{\mathrm{d}\log t}.  
    \label{eq:log_der_def}
\end{equation}
This logarithmic derivative quantifies the ``local'' algebraic decay of the function $f(t)$.  
The plot of $\mathrm{d}_{\mathrm{log}} \lambda^2C_{c, \mathrm{memo}}(t)$ is provided in Fig.~\ref{fig:crit_memo}(b). For $X_0$ finite, the curves exhibit a bump indicating the crossover, with the bumps occurring at increasingly longer times upon increasing $X_0$. At long times, the asymptotic value $-1/2$ of the logarithmic derivative corresponds to the algebraic decay presented in Fig.~\ref{fig:crit_memo}(a); at  short times, instead, the derivative takes the value $3/2$ which can be inferred from \cref{C_c_mem} in conjunction with \cref{mem_small_t}. 
This corresponds to the  initial rapid growth of $\lambda^2 C_{c, \mathrm{memo}}^{(2)}(t)$ observed in Fig.~\ref{fig:noncrit_memo}(a). 

Finally, we note that a similar long-time algebraic  behavior of $\lambda^2 C_{c, \mathrm{memo}}^{(2)}(t)$ with a $d$-dependent exponent, exhibiting a crossover from the bulk to the wall asymptotic, is expected for any $d \geq 1$.

\subsection{Power Spectral Density}
\begin{figure}
    \centering
    \begin{tabular}{cc}
        \includegraphics[width=0.49\linewidth]{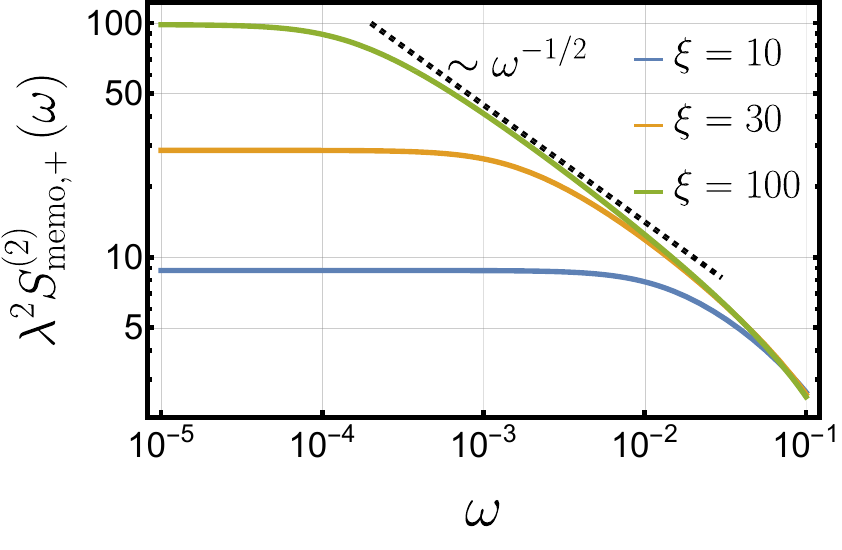}&
        \includegraphics[width=0.47\linewidth]{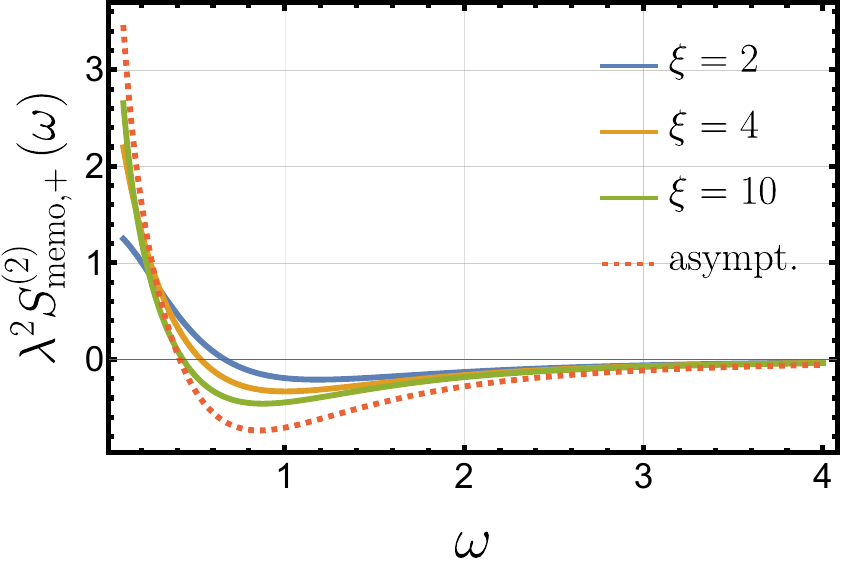}
        \\[-2mm]
         (a) & (b) 
    \end{tabular}
\caption{Bulk term of the memory-induced contribution $\lambda^2 S_{\mathrm{memo}, +}^{(2)}(\omega)$ to the power spectral density as a function of $\omega$, for various values of the correlation length $\xi$. (a) For large values of $\xi$, one observes $\lambda^2S^{(2)}_{\mathrm{memo}, +}(\omega) \propto \omega^{-1/2}$ (dotted line) at small but finite values of $\omega$. 
(b) At larger values of $\omega$,  instead, $\lambda^2S^{(2)}_{\mathrm{memo}, +}(\omega)$ shows a non-monotonic behavior and, in fact, after becoming negative, it approaches zero from below as $\omega \to\infty$. The various curves are compared with the one obtained at the critical point for strong confinement, see \cref{Spl_T0} (dashed line). The solid lines in both panels were obtained with $\lambda^2/(2 \kappa) = \ell_T = \sqrt{D/\omega_0} =1$. 
}
\label{fig:PSD1}
\end{figure}

In the context of the experimental analysis of the dynamics of tracer particles in fluid media, it is natural to consider the power spectral density (PSD) $S(\omega)$ \cite{Furst_book} 
of the particle position $X(t)$,
which is defined as the Fourier transform of $C_c(t)$:
            \begin{equation}
                S(\omega) =  \int \!\!\mathrm{d}t\, C_c(t)e^{i \omega t} = S^{(0)}(\omega) + \lambda^2 S^{(2)}(\omega) + \mathcal{O}(\lambda^4).
                \label{PSD_def}
            \end{equation}
In the decoupled case $\lambda=0$,  the zeroth-order term $S^{(0)}(\omega)$ corresponds to the PSD of the Ornstein-Uhlenbeck process
            \begin{equation}
                S^{(0)}(\omega) = \frac{\ell_T^2}{ \omega_0} \frac{2 \omega_0^2}{\omega_0^2 + \omega^2},
            \end{equation}
while the second-order term is given by
		\begin{equation}
			\lambda^2 S^{(2)}(\omega) =  \lambda^2 S_{\mathrm{C}}^{(2)}(\omega) + \lambda^2 S_{\mathrm{memo}}^{(2)}(\omega);
   \label{PSD}    
		\end{equation}
in this expression we distinguish, as done in \cref{subsec:corr}, the Casimir contribution
    \begin{equation}
        \lambda^2 S_{\mathrm{C}}^{(2)}(\omega) = - \frac{\lambda^2 \ell_T^2}{\kappa \omega_0\xi}\frac{4 \omega_0^4}{(\omega_0^2 + \omega^2)^2}e^{-\frac{2 X_0}{\xi} + \frac{2 T}{\kappa \xi^2}}
        \label{eq:S_Cas}
    \end{equation}
from the memory-induced contribution 
\begin{align}
        \lambda^2 & S_{\mathrm{memo}}^{(2)}(\omega) \n  \\ &=\frac{\lambda^2\ell_T^2}{\kappa \omega_0 \xi}\frac{2 \omega_0^4}{(\omega_0^2 + \omega^2)^2}  \int_0^{\infty}\!\!\mathrm{d}u\, \mathcal{F}(u)\left[  \cos\left(\frac{u\omega}{\omega_0}\right)\left(1 - \frac{\omega^2}{\omega_0^2}\right) -2\sin\left(\frac{u\omega}{\omega_0}\right)\frac{\omega}{\omega_0}\right] 
        \label{psd_mem}
    \end{align}
to the PSD, which is expressed in terms of $\mathcal{F}(u)$, given by Eq.~\eqref{dimless_memo}. 
The details of the derivation of these expressions are provided in \ref{PSD_app}.
Similarly to the case of the two-time correlator in Eq.~\eqref{2p_OU_renorm}, we rewrite the contribution to the PSD which is independent of the memory as
\begin{equation}
    S^{(0)}(\omega) + \lambda^2S^{(0)}_{\mathrm{C}}(\omega) = \frac{\ell_T'^2}{ \omega_0'}\frac{2 \omega_0'^2}{\omega_0'^2 + \omega^2} + \mathcal{O}(\lambda^4).
 \end{equation}
This equation expresses the PSD of the Ornstein-Uhlenbeck process with the renormalized trap strength $\kappa'$ given by Eq.~\eqref{kappa_renorm}, which renormalizes both the thermal lengthscale $\ell_T'$ and the relaxation rate $\omega_0'$, introduced after \cref{kappa_renorm} and before \cref{2p_OU_renorm}, respectively.
In order to investigate the properties of the memory-induced contribution to the PSD in Eq.~\eqref{psd_mem}, we express it as
    \begin{equation}
        \lambda^2 S_{\mathrm{memo}}^{(2)}(\omega) = \lambda^2 S_{\mathrm{memo}, +}^{(2)}(\omega) + \lambda^2 S_{\mathrm{memo}, - }^{(2)}(\omega),
        \label{psd_mem_decomp}
    \end{equation}
which follows from introducing in Eq.~\eqref{psd_mem} the decomposition of $\mathcal{F}(u)$ as the sum of the bulk term $\mathcal{F}^+(u)$ and the wall term $\mathcal{F}^-(u)$ according to \cref{dimless_mem_decomp}. It turns out that at the critical point $r=0$, both  $\lambda ^2S_{\mathrm{memo}, +}^{(2)}(\omega)$ and $\lambda ^2S_{\mathrm{memo}, -}^{(2)}(\omega)$ 
develop a singularity $\propto \omega^{-1/2}$ at small $\omega \to 0$. 
This asymptotic behavior is observed also for small but finite values of $\omega$ in the case of large but finite values of $\xi$ (i.e., for $r\neq 0$), as indicated by a dashed line in Fig.~\ref{fig:PSD1}(a), in which the bulk term $\lambda^2 S_{\mathrm{memo}, +}^{(2)}(\omega)$ of the memory-induced contribution to the PSD is plotted for $\lambda^2/(2\kappa) = \ell_T = \sqrt{D/\omega_0} = 1$.
To gain more insight into the behavior of $\lambda^2S_{\mathrm{memo}, \pm}^{(2)}$, we report here their expressions at the critical point and at the lowest order in 
$\ell_T$ (corresponding to the limit of strong confinement $\ell_T\to 0$), i.e., considering only the  dependence on $\ell_T$ of the prefactor $\ell_T^2$. 
In this limit, the bulk term of the memory-induced contribution is
\begin{align}
    \lambda^2 S^{(2)}_{\mathrm{memo}, +}(\omega) = \frac{\lambda^2 \ell_T^2}{2 \kappa \omega_0}\sqrt{\frac{2\omega_0}{D}}\frac{ \omega_0^4}{(\omega_0^2 + \omega^2)^2} \left(\frac{\omega}{\omega_0}\right)^{-1/2}\left[1 - 2\left(\frac{\omega}{\omega_0}\right)  - \left(\frac{\omega}{\omega_0}\right)^{2}\right],
    \label{Spl_T0} 
\end{align} 
and the wall term is
\begin{align}
    &\lambda^2 S^{(2)}_{\mathrm{memo}, -}(\omega)  = \frac{\lambda^2\ell_T^2}{2\kappa \omega_0}\sqrt{\frac{2\omega_0}{D}}\frac{ \omega_0^4}{(\omega_0^2 + \omega^2)^2}  \left(\frac{\omega}{\omega_0}\right)^{-1/2} \sqrt{2}e^{-\sqrt{\frac{2 \omega}{D}}X_0} 
    \times  \label{Smin_T0} \\ & \qquad \left\{\left[1 - \left(\frac{\omega}{\omega_0}\right)^{2}\right]\cos\left( \frac{\pi}{4} + \sqrt{\frac{2 \omega}{D}}X_0 \right) - 2\left(\frac{\omega}{\omega_0}\right) \sin\left( \frac{\pi}{4} + \sqrt{\frac{2 \omega}{D}}X_0 \right) \right\}. \n
\end{align}
Recall that the wall term is due to a mirror-image particle located at position $-X$.
We note that Eqs.~\eqref{Spl_T0} and \eqref{Smin_T0} coincide at $X_0 = 0$ because then the positions of the particle and its mirror image are the same.
Moreover, they exhibit the same behavior at small $\omega$, i.e., the prefactors of the singularity  $\propto \omega^{-1/2}$ are the same in both cases, irrespective of the value of $X_0$. 
This corresponds to the fact that, at long times, the presence of the wall doubles
the correlation of the particle position, which corresponds to the crossover of the correlation function presented in
\cref{fig:crit_memo}(a). 
Finally, we note that the wall term $\lambda^2 S^{(2)}_{\mathrm{memo}, -}(\omega)$ in Eq.~\eqref{Smin_T0} vanishes, upon increasing $\omega$, more rapidly than the bulk term $\lambda^2 S^{(2)}_{\mathrm{memo}, +}(\omega)$ in Eq.~\eqref{Spl_T0}, with an exponential decay $\propto \exp(-\sqrt{2 \omega/D}X_0)$ instead of the algebraic one observed in \cref{Spl_T0}. 
We recall that, at short times, the correlations due to the presence of the wall are extremely small, see \cref{mem_large_t}, which explains the large-$\omega$ behavior of the PSD discussed above.  

\begin{figure}
    \centering
    \begin{tabular}{cc}
        \includegraphics[width=0.47\linewidth]{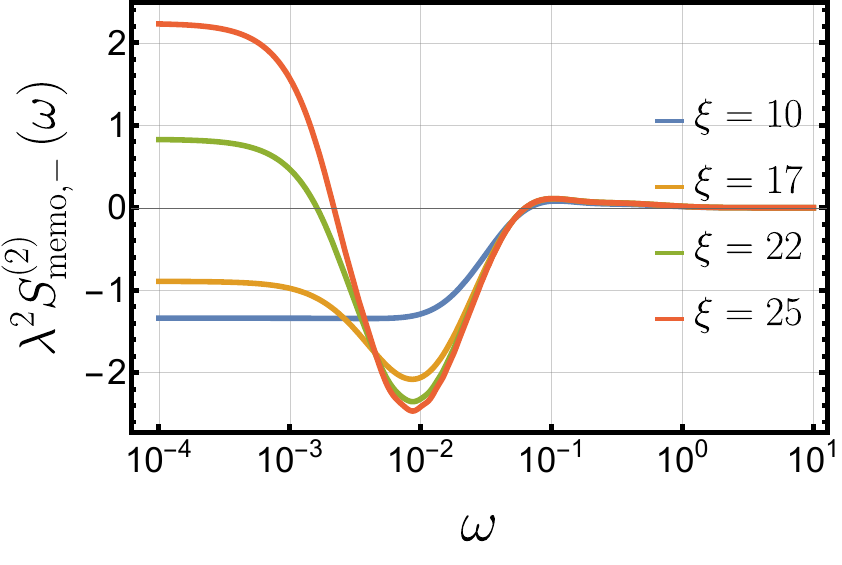}&
        \includegraphics[width=0.47\linewidth]{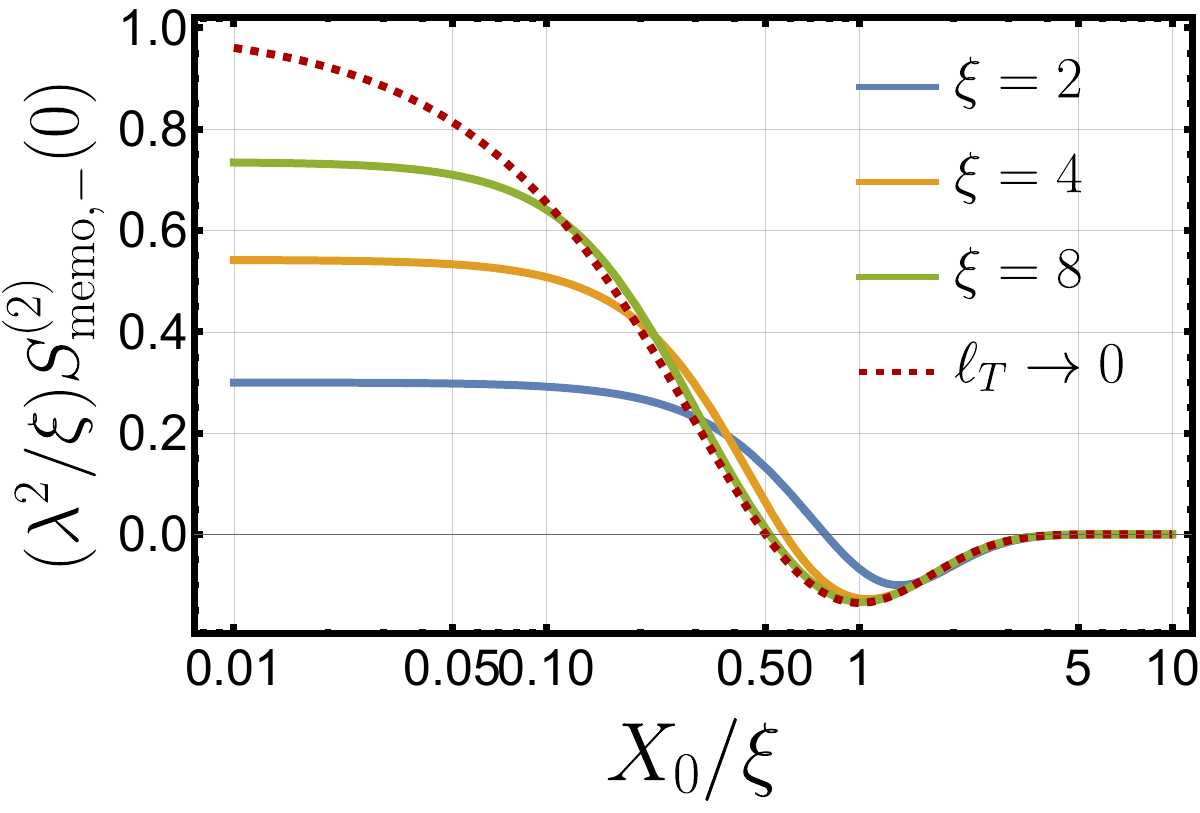}
        \\[-2mm]
         (a) & (b) 
    \end{tabular}
\caption{Wall term of the memory-induced contribution $\lambda^2 S_{\mathrm{memo}, -}^{(2)}(\omega)$ to the power spectral density for various values of the correlation length $\xi$, as a function of (a) $\omega$ with fixed $X_0=10$ or (b) 
$X_0$ with $\omega=0$. Note that  $\lambda^2 S_{\mathrm{memo}, -}^{(2)}(\omega)$ is generically a non-monotonic function of its variables and, for a fixed value of $X_0$, it attains a finite value for $\omega \to 0$, the sign of which 
depends on $\xi$, see panel (a). The dashed line in panel (b) corresponds to $(\lambda^2/\xi)S_{\mathrm{memo}, -}^{(2)}(\omega= 0)$ in the limit of strong confinement, given by Eq.~\eqref{S0_scaling}, which changes sign at $X_0/\xi =0.5$ and attains its minimal value at $X_0/\xi =1$. 
The solid lines in both panels are plotted for $\lambda^2/(2 \kappa) = \ell_T = \sqrt{D/\omega_0} =1$.}
\label{fig:PSD2}
\end{figure}

We note that, from the definition of the power spectral density in Eq.~\eqref{PSD_def}, it follows that 
\begin{equation}
    \int\!\!\frac{\mathrm{d} \omega}{2\pi} \, S(\omega) = C_c(0).
    \label{eq:int-sum}
\end{equation}
It can be shown by a direct calculation that the Casimir contribution to the PSD given by \cref{eq:S_Cas} satisfies
\begin{equation}
    \int \frac{\mathrm{d}\omega}{2 \pi}\lambda^2S_{\mathrm{C}}(\omega) = \lambda^2C_c^{(2)}(0),
\end{equation}
with $\lambda^2C_c^{(2)}(0)$ given by Eqs.~\eqref{C_c_2} and \eqref{C_c_C}, which implies, together with Eq.~\eqref{eq:int-sum}, that the integral over $\omega$ of the memory-induced contribution 
$\lambda^2 S^{(2)}_{\mathrm{memo}}(\omega)$ to the PSD in Eq.~\eqref{psd_mem} vanishes. 
This is actually expected, because it follows from the fact that the equal-time correlation of the particle position in the steady state is the equilibrium one, in which only the Casimir force $\mathbf{F}_{\mathrm{C}}$ appears.
Moreover, since this holds for any value of $X_0$, the wall and the bulk terms have to independently satisfy
\begin{equation}
    \int \frac{\mathrm{d}\omega}{2\pi} \lambda^2 S_{\mathrm{memo}, \pm}^{(2)}(\omega) = 0,
    \label{psd_0int_rel}
\end{equation}
as one can verify in the limit of strong confinement by a direct integration of  Eqs.~\eqref{Spl_T0} and \eqref{Smin_T0}.
Equation \eqref{psd_0int_rel} implies that $\lambda^2S_{\mathrm{memo}, \pm}^{(2)}(\omega)$ is a non-monotonic function of $\omega$, which takes both positive and negative values. This non-monotonic behavior is clearly visible in Figs.~\ref{fig:PSD1}(b) and~\ref{fig:PSD2}(a), in which $\lambda^2S_{\mathrm{memo}, +}(\omega)$ and $\lambda^2S_{\mathrm{memo}, -}(\omega)$ are plotted, respectively. 
Finally, we investigate the behavior of $\lambda^2 S_{\mathrm{memo}, \pm}^{(2)}(\omega)$ at $\omega = 0$, which is given by
\begin{align}
        \lambda^2 & S_{\mathrm{memo}, \pm}^{(2)}(\omega=0) =\frac{2\lambda^2 \ell_T^2}{\kappa \omega_0 \xi}  \int_0^{\infty}\!\!\mathrm{d}u\,\mathcal{F}^{\pm}(u).
        \label{eq:S0}
    \end{align}
It quantifies the net effect of the bulk and wall terms on the correlations of the particle position. 
For a finite  correlation length $\xi$, $\lambda^2 S_{\mathrm{memo}, +}^{(2)}(\omega = 0)$ has a finite positive value which grows linearly upon increasing $\xi$. 
As observed in \cref{fig:PSD2}(a), the wall term $\lambda^2 S_{\mathrm{memo}, -}^{(2)}(\omega = 0)$ may take both negative and positive values at $\omega=0$. To further examine this, we note that, in the limit of strong confinement $\ell_T \to 0$,  
\begin{equation}
    \lambda^2 S^{(2)}_{\mathrm{memo}, -}(\omega=0) = \frac{\lambda^2 \ell_T^2}{2 \kappa}\frac{\xi}{D}\left(1 - \frac{2X_0}{\xi}\right)e^{-\frac{2X0}{\xi}}.
    \label{S0_scaling}
\end{equation} 
Hence, in this limit, it is positive for $\xi > 2 X_0$ and negative for $\xi < 2 X_0$. Additionally, we plot $(\lambda^2/\xi)S_{\mathrm{memo}, -}^{(2)}(0)$  in Fig.~\ref{fig:PSD2}(b) for $X_0=10$ and $\lambda^2/(2 \kappa) = \ell_T = \sqrt{D/\omega_0} =1$ together with the asymptotic curve given by \cref{S0_scaling} indicated by a dashed line. 
The physical interpretation of \cref{S0_scaling} and \cref{fig:PSD2}(b)  is the following: in the limit of strong confinement, at the lowest order in $\lambda$, the particle localized at $X\simeq X_0$ affects the field in its vicinity. Such a perturbed portion of the medium is correlated over the distance given by the correlation length $\xi$. For $2X_0 < \xi$,  the correlations of the particle increase due to the presence of the wall (with positive $\lambda^2S_{\mathrm{memo}, - }(0)$), which stems from the effective interaction with the \textit{mirror image particle} trapped at $-X_0$. For $2X_0 > \xi$, the particle and its mirror image are uncorrelated. The only effect of the presence of the wall is that the field does not fill the entire space and there is less medium that could ``store'' the memory about the previous motion of the particle. Hence, the memory-induced correlations are suppressed and $\lambda^2S_{\mathrm{memo}, -}(0)$ is negative. Taking into account the thermal effects, the particle is effectively delocalized according to a Gaussian distribution and is spread over a typical distance given by $\ell_T$ defined in \cref{eq:lt_def}, and the plots in \cref{fig:PSD2}(b) flatten. 
Finally, we note that, independently of the value of $X_0/\xi$, the presence of the memory enhances correlations of the particle position, i.e., $\lambda^2 S^{(2)}_{\mathrm{memo}}(0)=\lambda^2 S^{(2)}_{\mathrm{memo}, +}(0) + \lambda^2 S^{(2)}_{\mathrm{memo},-}(0)$ is always positive.  

\section{Conclusions}
\label{sec_conclusions} 
We presented an analytical study of the equilibrium properties and of the effective dynamics of a tracer particle in a solvent, linearly coupled to a fluctuating Gaussian field with relaxational dissipative dynamics and subject to harmonic trapping.  The field has a tunable correlation length $\xi$, is subject to Dirichlet boundary conditions at the surface of a planar wall at a distance $X_0$ from the minimum of the trap, and fills the space only on one side of the wall. 
Both the particle and the field are in contact with a thermal bath with temperature $T$. In the present study, we assume $X_0$ to be much larger than the (effective) size of the particle and the typical lengthscale $\ell_T$ 
of the thermal fluctuations of the particle position inside the trap. 

First,  we derived the equilibrium distribution of the position $\mathbf{X}$ of the colloidal particle. We showed that, in addition to the harmonic potential of the optical trap, the Gibbs-Boltzmann probability distribution of $\mathbf{X}$ 
features an effective potential describing the interaction between the wall and the particle, mediated by the fluctuating field. The resulting fluctuation-induced force, which we called the Casimir-like force (due to its similarity to the forces acting on bodies imposing boundary conditions on fluctuations on their surfaces), turned out to be repulsive, i.e., it pushes the particle away from the wall. We introduced a scaling function for this
force in \cref{eq:Cas_scaling}, and showed that in the limit of a small particle  and at the critical point $\xi \to \infty$, the value of the force scales  $\sim X_1^{-(d-1)}$ upon increasing its distance $X_1$ from the wall, where $d$ is the dimensionality of the system.  
We demonstrated that, in the lowest quadratic order in $\lambda$, in $d=1$, and in the limit of a point-like particle, the Casimir-like force not only shifts  the average position of the particle $\langle X \rangle$, but also influences the variance $\langle X^2 \rangle - \langle X \rangle^2$ by renormalizing the strength of the optical trap, see Eq.~\eqref{kappa_renorm}, such that the thermal fluctuations of the particle position are actually reduced. At the critical point, the shift of the average position  $\langle X \rangle$ is maximal, whereas the effects on the variance vanish. 

We presented the effective dynamics of the particle in Eq.~\eqref{eff_dyn2}, which is expressed by a nonlinear non-Markovian Langevin equation featuring the force due to the trap, the fluctuation-induced force, a non-instantaneous friction, and an additional field-induced noise term.  The friction, which is not invariant under spatial translations, turned out to be related to the correlations of the field-induced noise via the extended fluctuation-dissipation relation reported in Eq.~\eqref{FDT_eff}, which we proved to ensure the invariance of the dynamics under time reversal.  
We discussed the adiabatic limit of the dynamics, i.e., the limit $D\to \infty$ of  the mobility $D$ of the field, with $\xi$ finite, in which the dynamics becomes Markovian, and the only effect of the coupling of the particle to the field is the emergence of the Casimir-like force. Finally, we noted that the linearization of the effective dynamics of the particle
features a multiplicative noise term and a linear memory kernel satisfying the usual fluctuation-dissipation relation \cite{Kubo_book}. 

Having discussed the properties of the exact effective dynamics of the tracer, we performed a perturbative calculation of the field-particle interaction for the connected two-time correlation function $C_c(t) = \langle X(t)X(0) \rangle - \langle X \rangle^2$  of the colloid position $X$, expanding in powers of  the coupling constant $\lambda$. 
For simplicity, we focused on the one-dimensional case $d=1$, assuming a point-like particle. 
The lowest-order correction, quadratic in $\lambda$, featured two terms: (i) a Casimir term $\lambda^2C^{(2)}_{c, \mathrm{C}}(t)$ which renders the change of the variance of the particle  position at $t = 0$.  The sum of this term and the correlator obtained in the decoupled case ($\lambda=0$) renders the two-point function of the Ornstein-Uhlenbeck process with the renormalized trapping strength mentioned above; (ii) a memory-induced term $\lambda^2C_{c, \mathrm{memo}}(t)$ which vanishes for $t=0$.  We expressed the latter in terms of the dimensionless memory $\mathcal{F}$, see \cref{C_c_mem}, and  demonstrated its dependence on the distance between the center of the optical trap and the wall in the case of the system at criticality. In particular, at short times, the effects due to the presence of the wall are negligible, whereas at longer times, the presence of the wall doubles the memory-induced correlations because of an effective interaction with a \textit{mirror image particle}  at position $-X$.

We discussed the behavior of $\lambda^2C_{c, \mathrm{memo}}(t)$ which, for finite values of $X_0$, displays a crossover from the asymptotic behavior corresponding to the bulk $X_0 \to \infty$ case, featuring a decay $\sim t^{-1/2}$ at long times, to that corresponding to the trap placed exactly at the wall, characterized by the same power-law, but with a prefactor 
twice as large as in the bulk.
We also discussed the properties of the power spectral density (PSD),  i.e., the Fourier transform of the two-time correlation of the particle position, focusing in particular on the non-monotonic behavior of the memory-induced contribution to  the PSD. Finally, we showed that the presence of the wall might enhance or diminish the correlations of the particle position, depending on the distance of the particle from the wall, on the value of the correlation length, and on the spatial extent of thermal fluctuations of the particle position, quantified by the PSD at zero frequency.

The results presented above motivate further investigations of the behavior of tracer particles in  correlated media in the presence of a wall. In fact, several open questions remain, including the impact of different forms of coupling between the field, the wall, and the particle -- such as symmetry-breaking boundary conditions at the wall or a quadratic coupling to the particle -- that may lead to attractive fluctuation-induced forces. Future work should also consider the role of the self-interaction for the field (e.g., $\propto \phi^4$) and possible hydrodynamic effects in the particle and the field dynamics.  In fact, they are key elements of more realistic models of the dynamics of near-critical media. Additionally, the regime in which the particle remains confined near the wall --- which requires accouting for a  direct wall-particle interactions --- warrants additional study. 
Finally, the present analysis can be extended to non-equilibrium settings, including scenarios where the particle is dragged by a trap moving at constant velocity, oscillating at a fixed frequency, or in active-matter systems in the presence of a wall.
    
\section*{Acknowledgments} We thank Sarah A.~M.~Loos, Marco Baiesi, Aljaz Godec and Davide Venturelli for constructive discussions. 

\appendix

\addtocontents{toc}{\fixappendix}

\section{Equilibrium distribution}

In this Appendix, we sketch the calculations yielding the equilibrium properties of the system.  In particular, in \ref{per}, the expression of the marginal equilibrium distribution of the particle position in Eq.~\eqref{pdf_eq} and the corresponding effective potential \eqref{V_eff} are derived. 
In \ref{app_pert_eq}, we turn our attention to the case of $d=1$ with a point-like particle \eqref{eq:kernel_point_like_lim}, and sketch the perturbative calculation of the lowest-order correction in $\lambda$  to $\langle X \rangle$ and $\langle X^2 \rangle - \langle X \rangle^2$, reported in \cref{1p_eq_1d} and \cref{2p_eq_1d} respectively.

\subsection{Effective Potential}
\label{per}

In equilibrium, the probability distribution describing the position $\mathbf{X}$ of the colloidal particle is given by the marginal of the Gibbs-Boltzmann distribution in Eq.~\eqref{eq:pdf_marg_def}.  
To trace out the field degrees of freedom therein, i.e., to perform the functional integration, we express the field-dependent part of the Hamiltonian in Eqs.~\eqref{eq:gaussian_hamiltonian_rec} and \eqref{eq:Hint} in a bilinear form
\begin{align}    
    \mathcal{H}_{\phi}[\phi] + &\mathcal{H}_{\mathrm{int}}[\phi, \mathbf{X}]\n  \\  &=  \int_{\Lambda} \mathrm{d}^d\mathbf{x}\int_{\Lambda} \mathrm{d}^d\mathbf{x}' \left\{ \frac{1}{2}\phi(\mathbf{x})\mathbf{\hat{A}}(\mathbf{x}, \mathbf{x}')\phi(\mathbf{x})\right\} - \lambda\int_{\Lambda} \mathrm{d}^d\mathbf{x} \, \phi(\mathbf{x}) V(\mathbf{x}-\mathbf{X})  ,
    \label{Ham_app}
\end{align}
with $\mathbf{\hat{A}}(\mathbf{x}, \mathbf{x}') = \delta^{(d)}(\mathbf{x} - \mathbf{x}')(r - \nabla^2 )$ and $\Lambda = \mathbb{R}_+\times\mathbb{R}^{d-1}$. The functional integration in \cref{eq:pdf_marg_def} is Gaussian, and hence, it can be evaluated
\begin{align}
    \int \mathcal{D'}\phi & e^{-\left[\mathcal{H}_{\phi}[\phi] + \mathcal{H}_{\mathrm{int}}[\phi, \mathbf{X}]\right]/T} \n \\ &\propto \exp\left[\frac{\lambda^2}{2T} \int_{\Lambda} \mathrm{d}^d\mathbf{x}\int_{\Lambda} \mathrm{d}^d\mathbf{x}' V(\mathbf{x} - \mathbf{X})\mathbf{\hat{A}^{-1}}(\mathbf{x}, \mathbf{x'})V(\mathbf{x}' - \mathbf{X} )  \right], 
\end{align}
where $\mathbf{\hat{A}^{-1}}$ is the inverse operator of $\mathbf{\hat{A}}$. These operators are not translationally invariant since they are defined on the space of fields vanishing at $x_1 = 0$. 
To determine $\mathbf{\hat{A}^{-1}(\mathbf{x}, \mathbf{x'})}$, we express the field in the basis of combinations of plane waves vanishing at the wall, i.e., we write
\begin{equation}
    \phi(\mathbf{x}) =  \int_{\Lambda} \frac{\mathrm{d}^{d} \mathbf{k}}{(2 \pi)^{d}}\, \phi_{\mathbf{k}} \, 2\sin(k_1x_1)e^{-i \mathbf{k_{\parallel}} \cdot\mathbf{x_{\parallel}}} ,
    \label{plane_wave}
\end{equation}
which satisfies
\begin{equation}
    \phi_{\mathbf{k}} = \int_{\Lambda} \mathrm{d}{}^d \mathbf{x}\,  \phi(\mathbf{x}) \,  2\sin (k_1x_1)e^{i \mathbf{k_{\parallel}} \cdot\mathbf{x_{\parallel}}}.
\end{equation}
Then, one can cast the operator $
\mathbf{\hat{A}(x, x')}$ in this basis 
\begin{equation}
    \mathbf{\tilde{A}(k, k')} = (r + k^2)(2 \pi)^{d}\delta(k_1 - k'_1) \delta^{(d-1)}\left(\mathbf{k}_{\parallel} + \mathbf{k'}_{\parallel}\right)  ,
\end{equation}
whose inverse is given by
\begin{equation}
    \mathbf{\tilde{A}^{-1}(k, k')} =  \frac{1}{r + k^2}(2 \pi)^d\delta(k_1 - k'_1)\delta^{(d-1)}\left(\mathbf{k}_{\parallel} + \mathbf{k'}_{\parallel}\right)   .
\end{equation}
Using the expansion in plane waves in Eq.~(\ref{plane_wave}), we obtain 
\begin{equation}
    \mathbf{\hat{A}^{-1}}(\mathbf{x}, \mathbf{x'}) = \int \frac{\mathrm{d}^{d}\mathbf{k}}{(2 \pi)^d}\frac{1}
{r + k^2}\left[e^{-i \mathbf{k} \cdot (\mathbf{x}-\mathbf{x'})} - e^{-i \mathbf{k} \cdot (\mathbf{x}-\mathbf{x'}_R)}\right] \equiv C_{\phi,\textrm{st}}(\mathbf{x}),
\end{equation}
with $C_{\phi,\textrm{st}}(\mathbf{x})$ defined in \cref{V_eff}. 
Performing the above integration yields the results reported in Eqs.~(\ref{pdf_eq}), (\ref{V_eff}) and (\ref{propagators}).

\subsection{Equilibrium perturbative calculation for the 1-point and 2-point functions}
\label{app_pert_eq}

In this Appendix, we consider the case of the point-like particle in Eq.~\eqref{eq:kernel_point_like_lim} in one spatial dimension $d=1$. We briefly describe the calculation of the lowest-order correction in $\lambda$ to the average and the variance of the position of this particle, which are reported in Eqs.~\eqref{1p_eq_1d} and \eqref{2p_eq_1d}, respectively.
Since $V_{\mathrm{eff}}(X_1)$ is $\mathcal{O}(\lambda^2)$, expanding the exponent in \cref{pdf_eq} in powers of $\lambda$ yields
\begin{equation}
    \exp\left\{   -\frac{\kappa (\mathbf{X} - \mathbf{X}_0)^2}{2T} - \frac{V_{\mathrm{eff}}(X_1)}{T} \right\}  = e^{-\frac{\kappa (\mathbf{X} - \mathbf{X}_0)^2}{2T} } \left(1 - \frac{V_{\mathrm{eff}}(X_1)}{T}  \right) + \mathcal{O}(\lambda^4).
\end{equation}
For the delta-like particle-field interactions considered here, the translationally invariant part of the static correlation function $C^D_{\phi,\rm{st}}(\mathbf{x},\mathbf{x}')$, i.e., the part which is a function of $(\mathbf{x} - \mathbf{x}')$ in  Eqs.~\eqref{V_eff}, \eqref{Dpropagators} and \eqref{propagators} is irrelevant, as it renders an $X_1$-independent constant. Hence,  one can focus on  
\begin{equation}
    \tilde V_{\mathrm{eff}}(X_1) = \frac{\lambda^2 }{2}C_{\phi,\textrm{st}}(2X_1).
\end{equation}
 Assuming \eqref{eq:X0_gg_lT},  integration over $\Lambda = \mathbb{R_+} $ can be approximated by the integration over~$\mathbb{R}$. In addition, the same assumption allows us to approximate the  argument  $|2X_1|/\xi$ of the exponential in the effective potential (see \cref{propagators}) by $2X_1/\xi$, since the contribution from $X_1<0$ will be negligibly small.  Accordingly, these approximations render the relevant integrals Gaussian, and a straightforward calculation for the normalization constant $\mathcal{N}_X$ in 
 \cref{pdf_eq} within this perturbative approach -leads to
\begin{equation}
    \mathcal{N}_X = 
        \sqrt{2 \pi }\ell_T\left(1 - \frac{\lambda^2  \xi}{4T} e^{-\frac{2 X_0}{\xi} + \frac{2T}{\kappa \xi^2}} \right) + \mathcal{O}(\lambda^4).
\end{equation}
From this,  averages of $X$ and $X^2$ are obtained from straightforward Gaussian integrals over $\mathbb{R}$, which are reported in Eqs.~\eqref{1p_eq_1d} and~\eqref{2p_eq_1d}, respectively.

\section{Adiabatic approximation}
\label{ad_app}

In this Appendix, we elaborate on the concept of the \textit{adiabatic limit}, in which the  effective dynamics of the colloid in \cref{eff_dyn1} becomes Markovian. For convenience, let us consider the case $d = 1$.
The approximation of instantaneous equilibration of the medium formally corresponds to taking the limit $D \to \infty$ while keeping $r$ (equivalently $\xi$) finite. It can be shown that, for a $D$-independent function $f(t)$,
\begin{equation}
    \int_{-\infty}^{t}\!\!\mathrm{d}t' \, R_{\phi}^D(x,y, t- t')f(t') \xrightarrow[D \to \infty]{} \frac{\xi}{2} 
    \left[ e^{-\frac{|x - y|}{\xi}} - e^{-\frac{|x + y|}{\xi}} \right] f(t),
\label{ab_app}
\end{equation}
where $R_{\phi}^D$ is the response function satisfying the Dirichlet boundary conditions at the wall, introduced in Eqs.~\eqref{response2} and \eqref{response1}.
From the property above, it follows that
\begin{equation}
    \int_{-\infty}^{t}\!\!\mathrm{d}t' \, \frac{1}{D}R_{\phi}^D(x, y, t- t')f(t') \xrightarrow[D \to \infty]{} 0.
    \label{ab_app2}
\end{equation}
Accordingly, from Eqs~\eqref{V_eff},~\eqref{propagators}, \eqref{Cas_def}, and (\ref{ab_app}),  the function $F$ introduced in the effective dynamics \eqref{eff_dyn1} and given by \cref{f_funct} (here we drop the vectorial notation, given that the force $\mathbf{F}$ has only one component $F$ in $d=1$)  satisfies 
\begin{equation}
    \int_{-\infty}^t \mathrm{d} t' F\left(X(t), X(t'), t - t'\right) \xrightarrow[D \to \infty]{} F_{\mathrm{C}}(X(t)),
\end{equation}
where $F_{\mathrm{C}}$ is the Casimir-like force defined in  \cref{Cas_def}.
Moreover, the field-induced noise $\Xi$ introduced in \cref{eff_dyn1} is given by
\begin{equation}
    \Xi(x, t) = -\lambda \int_{0}^\infty \mathrm{d}z \int_{0}^\infty \mathrm{d}z'\, V'(z - x)\int_{-\infty}^t\mathrm{d}t' \, \frac{1}{D}R_{\phi}^D(z, z', t-t')\xi(z', t')
    \label{Xi_def}
\end{equation}
and, according to Eq.~(\ref{ab_app2}), satisfies
\begin{equation}
    \Xi \left( X(t), t\right) \xrightarrow[D \to \infty]{} 0,
\end{equation}
where $\Xi$ indicates the only component of $\bm{\Xi}$ in $d=1$.
 An analogous argument holds in higher spatial dimensionality $d>1$.
Accordingly, in the adiabatic approximation, the effective dynamics is Markovian and given by \cref{eff_dyn_markov}.

\section{Time reversal symmetry}
\label{app_t_reversal}

In this Appendix, we show that the effective dynamics in Eq.~\eqref{eff_dyn1} with the noise term $\bm{\eta}$ satisfying Eq.~\eqref{eq:part_noise},  the field-induced noise $\bm{\Xi}$, characterized by the variance  \eqref{eq:eff_dyn_noise} and expressed in terms of the matrix elements of $\mathbb{G}$, and  the nonlinear  force $\bm{F}$ is indeed invariant under time-reversal, provided that the latter two satisfy the relation in Eq.~\eqref{FDT_eff}. 
A similar argument was presented in Ref.~\cite{Basu_2022} in the case of the particle in the bulk. 
The argument will be provided for the case $d=1$, but it readily generalizes to higher dimensions. For clarity of presentation, we omit the vectorial notation, understanding that in $d=1$ all quantities under investigation are scalar.

We start by noticing that Eq.~\eqref{FDT_eff} makes it convenient to introduce a function $\psi(x, y, t)$ given by 
    \begin{equation}
        \psi(x, y, t) = \int_t^{\infty}\!\!\mathrm{d}t'\, F(x, y, t'),
    \end{equation}
which satisfies 
    \begin{align}
            &F(x, y, t) = -\partial_t \psi(x, y, t), 
            \label{psi_rel_1} \\ 
            &G(x, y, t) = \partial_y \psi(x, y, t).
            \label{psi_rel_2}
    \end{align}
Moreover, we note that
        \begin{equation}
            F_{\mathrm{C}}(x) =  -\frac{\mathrm{d}}{\mathrm{d}x}V_{\mathrm{eff}}(x) = \int_{-\infty}^{t}\!\!\mathrm{d}t'\, F(x, x, t-t') = \psi(x, x, 0).
        \label{psi_Cas}
        \end{equation}  
Additionally, the following symmetries of $G(x, y, t)$ will prove useful in what follows:
\begin{align}
    G(x, y, t) =  G(x, y, |t|), 
    \label{G_sym0} \\ 
     G(y, x, t) =  G(x, y, t).
     \label{G_sym}
\end{align}
In order to investigate the possible time-reversal symmetry of the dynamics, we introduce its path-integral representation in terms of the Martin-Siggia-Rose \cite{Martin_Siggia_Rose, De_Dominicis_1976, Janssen_1976} 
action 
		\begin{equation*}
			S[x(t), p(t)] = S_0[x(t), p(t)] + S_{\mathrm{int}}[x(t), p(t)],
		\end{equation*}
where $S_0$ is the action without the field (Ornstein-Uhlenbeck process with $\lambda=0$) and $S_{\mathrm{int}}$ is the term stemming from the particle-field interaction. The former is given by
            \begin{align}    
                S_0& [x(t), p(t)] \n \\  &= \frac{\kappa\left(x_{- \infty} -X_0\right)^2}{2T} - i\int \mathrm{d}t\, p(t) \left[\gamma_0 \dot x(t) + \kappa (x(t) - X_0) \right] + \gamma_0 T \int \mathrm{d} t\, p^2(t),
            \end{align}    
            where $x_{-\infty} = x(t \to -\infty)$ is the initial condition drawn from the Gibbs-Boltzmann distribution, and $p(t)$ is the so-called response field. As demonstrated in Ref.~\cite{Demery_2011}, $S_{\rm int}$ is given by
		\begin{align}
			S_{\mathrm{int}}[x(t), p(t)] = \frac{V_{\mathrm{eff}}\left(x_{- \infty}\right)}{T} &+ i \int \mathrm{d}t \mathrm{d}t' \, \theta \left( t - t' \right) p(t) F\left( x(t), x(t'), t-t' \right) \n  \\ & + \frac{T}{2} \int \mathrm{d}t\mathrm{d}t' \, p(t) G\left( x(t), x(t'), t-t' \right) p(t').
            \label{eq:app_action_og}
		\end{align}
            We can now use the method presented in Ref.~\cite{Aron_2010} to show that the relations (\ref{psi_rel_1}) and \eqref{psi_rel_2} accompanied by Eq.~(\ref{psi_Cas}) guarantee time-reversal symmetry of the stationary process described by the considered dynamics, i.e., that it is an equilibrium process. For a given trajectory $\{x(t), p(t) \}$, its time-reversed counterpart $\{\bar{x}(t), \bar{p}(t) \}$ is given by  \cite{Aron_2010}
		\begin{align}
				&x(t) \mapsto \bar{x}(t) = x(-t), \\
				&p(t) \mapsto \bar{p}(t) = p(-t) - \frac{i}{T}\dot{x}(-t),
		\end{align}
  where the dot above a function denotes the derivative with respect to the argument. In the case of equilibrium processes, the corresponding action satisfies $S[x, p] = S[\bar{x}, \bar{p}]$ which is the property that we prove now.
  It can be easily shown that $S_0[\bar{x}, \bar{p}] =  S_0[x, p]$. The proof of the fact that $S_{\mathrm{int}}[\bar{x}, \bar{p}] =  S_{\mathrm{int}}[x, p]$ is a bit more involved and we present it below. 	
Using the symmetries of $G(x, y, t)$ reported in Eqs.~(\ref{G_sym0}) and (\ref{G_sym}), we can write the time-reversed action in the form
\begin{align}
    S_{\mathrm{int}}&[\bar{x}(t), \bar{p}(t)] =   \frac{V_{\mathrm{eff}}(x_{\infty})}{T} \n \\ &+ \frac{1}{T} \int \mathrm{d}t \mathrm{d}t' \, \theta(t' - t)\dot x(t) \left[F(x(t), x(t'), t'-t) - \dot x(t') G(x(t), x(t'), t'-t) \right] \n \\ &+ i \int \mathrm{d}t \mathrm{d}t' \, \theta(t' - t)p(t) \left[F(x(t), x(t'), t'-t) - \dot x(t') G(x(t), x(t'), t'-t) \right]  \n \\ &- i \int \mathrm{d}t \mathrm{d}t' \, \theta(t - t')p(t)  G(x(t), x(t'), t-t') \dot x(t')\n \\  &+ \frac{T}{2} \int \mathrm{d}t \mathrm{d}t'\, p(t) G(x(t), x(t'), t-t')p(t').  
    \label{eq:TRA-1}
\end{align}
We note that the last term in the time-reversed action coincides with the last term in the original action \eqref{eq:app_action_og}. From the relations given in  Eqs.~(\ref{psi_rel_1}) and (\ref{psi_rel_2}), one obtains
\begin{align}
        -\frac{\mathrm{d}}{\mathrm{d}t'}\psi(x(t), x(t'), t' - t) = F(x(t), x(t'), t' - t) - \dot x(t') G(x(t), x(t'), t' - t), \\ 
        \frac{\mathrm{d}}{\mathrm{d}t'} \psi(x(t), x(t'), t - t') = F(x(t), x(t'), t - t') + \dot x(t') G(x(t), x(t'), t - t').
\end{align}
Accordingly, the time-reversed action in Eq.~\eqref{eq:TRA-1} can be written as 
\begin{align}
    S_{\mathrm{int}}[\bar{x}(t), \bar{p}(t)] &=   \frac{V_{\mathrm{eff}}(x_{\infty})}{T} - \frac{1}{T} \int \mathrm{d}t \, \dot x(t) \int_{t}^{\infty} \!\!\mathrm{d}t'\,\frac{\mathrm{d}}{\mathrm{d} t'} \left[\psi(x(t), x(t'), t'-t)\right] \n \\ & - i \int \mathrm{d}t \,  p(t) \int_{t}^{\infty}\!\! \mathrm{d}t'\frac{\mathrm{d}}{\mathrm{d} t'} \left[\psi(x(t), x(t'), t'-t)\right] \n \\ & - i \int \mathrm{d}t  \, p(t) \int_{-\infty}^{t} \!\!\mathrm{d}t'\, \frac{\mathrm{d}}{\mathrm{d} t'}  \left[\psi(x(t), x(t'), t-t')\right]  \label{1111}  \\ &+i \int \mathrm{d}t \,\mathrm{d}t' \, \theta(t - t')p(t) F(x(t), x(t'), t-t') \n \\ &  + \frac{T}{2  } \int \mathrm{d}t \, \mathrm{d}t' \, p(t) G(x(t), x(t'), t-t')p(t'). \n
\end{align}
We notice that $\psi(x, y, t) \xrightarrow[t \to \infty]{} 0$ and that the two integrals in the second and 
the third line of \cref{1111} cancel each other out. Moreover, the last two terms coincide with the last two terms of the forward-time action in Eq.~\eqref{eq:app_action_og}. Hence, in order to prove time-reversal symmetry, we have to show that the first term in  Eq.~\eqref{eq:app_action_og} is equal to the two terms in the first line Eq.~\eqref{1111}. We note that, due to Eq.~\eqref{psi_Cas},
\begin{align}    
    \frac{V_{\mathrm{eff}}(x_{\infty})}{T} & - \frac{1}{T} \int \mathrm{d}t \dot x(t) \int_{t}^{\infty} \mathrm{d}t'\frac{\mathrm{d}}{\mathrm{d} t'} \left[\psi(x(t), x(t'), t'-t)\right] \n \\  &= \frac{V_{\mathrm{eff}}(x_{\infty})}{T} + \frac{1}{T} \int \mathrm{d}t \dot x(t) \psi(x(t), x(t), 0) \n \\   &= \frac{V_{\mathrm{eff}}(x_{\infty})}{T} - \frac{1}{T} \int \mathrm{d}t \frac{\mathrm{d} x}{\mathrm{d}t} \frac{\mathrm{d}V_{\mathrm{eff}}(x(t)) }{\mathrm{d}x(t)} \n \\ &= \frac{V_{\mathrm{eff}}(x_{\infty})}{T} - \frac{1}{T} \int_{x_{-\infty}}^{x_{\infty}} \mathrm{d}x  \frac{\mathrm{d}V_{\mathrm{eff}}(x) }{\mathrm{d}x} = \frac{V_{\mathrm{eff}}(x_{-\infty})}{T}
\end{align}    
which concludes the proof of time-reversal symmetry of the action of the process.

\section{Perturbative expansion}
\label{app_pert_exp}

In this Appendix, we solve the dynamics given by Eqs.~(\ref{dyn_both1}) and (\ref{dyn_both2}) in the case of a one-dimensional system with a point-like particle, see Eq.~\eqref{eq:kernel_point_like_lim}. 
First, we state the formal solutions of the dynamics within the perturbative expansion  \eqref{pert_exp_def_field} and \eqref{pert_exp_def_part}, and report properties of the Ornstein-Uhlenbeck process relevant for further calculations. Then, in \ref{1point_app}, we demonstrate that evaluating the lowest-order correction to the average particle position obtained from the equations of motion yields the same result as the one obtained from the Gibbs-Boltzmann distribution, reported in \cref{2p_eq_1d}. Finally, we derive the lowest-order correction to the connected two-point function, reported in Eqs.~\eqref{C_c_2}, \eqref{C_c_C}, and \eqref{C_c_mem}.

Applying the expansion in powers of $\lambda$ to $\phi$ and $X$, see Eqs.~\eqref{pert_exp_def_field} and \eqref{pert_exp_def_part} respectively,  to the dynamics~(\ref{dyn_both1}) and (\ref{dyn_both2}), and solving it term-by-term for the field configuration yields
\begin{align}   
    &\phi^{(0)}(x, t) = \int_{-\infty}^{t}\mathrm{d}t' \int_{0}^{\infty}\mathrm{d}x' \, \frac{1}{D} R_{\phi}^D(x, x', t-t')\xi(x', t'), \label{phi_0} \\
    &\lambda \phi^{(1)}(x, t) = \lambda \int_{-\infty}^{t}\mathrm{d}t' \int_{0}^{\infty}\mathrm{d}x'\, R_{\phi}^D(x, x', t-t')V(x' - X^{(0)}(t')),
    \label{phi_1} 
\end{align}
where the $R_{\phi}^D$ is defined in Eqs.~\eqref{response2} and \eqref{response1}. Solving for the particle position and using Eq.~\eqref{eq:X0_gg_lT} renders
\begin{align}
         &X^{(0)}(t) = X_0 + \gamma_0^{-1} \int_{-\infty}^t \!\!\mathrm{d} t'\, e^{-\omega_0 (t - t')}\eta(t') =  X_0 + \tilde X(t),\label{X_0} \\
         \lambda &X^{(1)}(t) = \gamma_0^{-1} \int_{-\infty}^t \!\!\mathrm{d} t'\, e^{-\omega_0 (t - t')}\Xi(X^{(0)}(t'), t'), \label{X_1} \\ 
        \lambda^2 &X^{(2)}(t) = \gamma_0^{-1} \int_{-\infty}^t \!\!\mathrm{d} t'\, e^{-\omega_0 (t - t')} \int_{-\infty}^{t'} \!\!\mathrm{d}t''F(X^{(0)}(t'), X^{(0)}(t''), t'- t'')  \n \\ 
          & \qquad \qquad \quad +\gamma_0^{-1} \int_{-\infty}^t \!\!\mathrm{d} t'\, e^{-\omega_0 (t - t')} \lambda X^{(1)}(t')  \partial_{X} \Xi(X^{(0)}(t'), t') \label{X_2},
\end{align}
where $\omega_0 $ is defined below \cref{C_c_0}. While deriving $X^{(1)}$, we used the expressions of $\phi^{(0)}$ and $\Xi$  in Eqs.~\eqref{phi_0}  and  \eqref{Xi_def}, respectively. 
Note that in Eq.~\eqref{X_0} we introduced $\tilde{X}(t)$, which is the Ornstein-Uhlenbeck process characterized by the transition probability 
\begin{equation}
    \mathcal{P}(x_1, t_1 | x_2, t_2) = \frac{1}{\sqrt{2 \pi \ell_T^2(1 - e^{-2\omega_0 (t_1 - t_2)})} }\exp\left\{ - \frac{1}{2}\frac{(x_1 - x_2e^{-\omega_0 (t_1 - t_2)})^2}{\ell_T^2\left(1 - e^{-2\omega_0 (t_1 - t_2)}\right)} \right\},
    \label{OU_pdf}
\end{equation}
where $\ell_T$ is the thermal lengthscale defined below \cref{eq:lt_def}. 
 For the sake of further derivations, using Eq.~(\ref{OU_pdf}), we evaluate the following quantities
\begin{equation}
    \langle e^{ik_1 X^{(0)}(t_1) + ik_2 X^{(0)}(t_2)}\rangle = \exp\left [{i(k_1  + k_2) X_0 - \frac{\ell_T^2}{2}\left(k_1^2 + 2k_1k_2e^{-\omega_0 |t_1 - t_2|} + k_2^2\right)}\right],
    \label{app_exp_avg1}
\end{equation}
and
\begin{align}    
    \langle & e^{ik_1 X^{(0)}(t_1) + ik_2 X^{(0)}(t_2)} \tilde{X}(t_3)\rangle \label{app_exp_avg2} \\  &=  i\ell_T^2\left(k_1e^{-\omega_0 |t_1 - t_3|} + k_2e^{-\omega_0 |t_2 - t_3|}\right)  \exp \left[{i(k_1  + k_2) X_0 - \frac{\ell_T^2}{2}\left(k_1^2 + 2k_1k_2e^{-\omega_0 |t_1 - t_2|} + k_2^2\right)}\right]. \n 
\end{align}
Additionally, we note that for a point-like particle \cref{f_funct} renders
\begin{equation}
    F(x, y, t) = -i\lambda^2  D \int\frac{\mathrm{d} k}{2 \pi}\,  k\,e^{-D(k^2+r)t}\left[e^{-ik(x - y)} - e^{-ik(x + y)} \right]
    \label{eq:app_pointlike_F}
\end{equation}
while
\begin{equation}
    G(x, y, t) = \lambda^2  \int \frac{\mathrm{d}k}{2\pi} \, \frac{k^2}{k^2 + r}e^{-D(k^2 + r)|t|}\left[e^{-ik(x - y)} + e^{-ik(x + y)} \right],
    \label{eq:app_pointlike_G}
\end{equation}
which follows from \cref{g_funct}, where in both cases we omit the indication of the component of the vector, as we are considering the case $d=1$.
We conclude the preliminary remarks of the Appendix by recalling that the equations of motion \eqref{dyn_both1} and \eqref{dyn_both2} are invariant under the transformation \eqref{parity}. Hence, $\langle X\rangle$ and $\langle X^2\rangle$ are  even functions of $\lambda$.

\subsection{The average position}
\label{1point_app}
The  lowest-order  correction to the average position of the particle is given by  
\begin{multline}
        \lambda^2 \langle  X^{(2)}(t) \rangle   = \frac{1}{\gamma_0}\int_{-\infty}^t\!\! \mathrm{d}t'\,e^{-\omega_0 (t -t')} \int_{-\infty}^{t'} \!\!\mathrm{d}t'' \, \left\langle F\left(X^{(0)}(t'), X^{(0)}(t'') , t' - t''\right) \right\rangle  + \\ \frac{T}{\gamma_0^2}\int_{-\infty}^t \!\!\mathrm{d}t'\,e^{-\omega_0 (t -t')} \int_{-\infty}^{t'}\!\! \mathrm{d}t''\, e^{-\omega_0(t' - t'')}\left\langle \partial_{X^{(0)}(t')}G\left(X^{(0)}(t'), X^{(0)}(t'') , t' - t''\right) \right\rangle,
        \label{1p_pert_app}
\end{multline}
 where we used Eq.~\eqref{eq:eff_dyn_noise} for the average of the field-induced noise. 
Making use of Eqs.~\eqref{app_exp_avg1} and \eqref{eq:app_pointlike_F} renders the first term in the sum on the right-hand side of \cref{1p_pert_app}, i.e.,
\begin{multline}
    \left\langle F\left( X^{(0)}(t'), X^{(0)}(t''),  t' - t''\right)\right\rangle  \\ 
    = \lambda^2  D \int\frac{\mathrm{d}k}{2 \pi}  e^{-D(k^2+r) (t' - t'') - \ell_T^2k^2\left(1 + e^{-\omega_0(t' - t'')}\right)} k\, \sin(2 X_0 k)
    \label{1p_pert_app_term_1}
\end{multline}
In a similar way, from Eqs.~\eqref{eq:app_pointlike_G} and \eqref{app_exp_avg1}, we obtain
\begin{multline}
    \left\langle \partial_{X^{(0)}(t')}G\left(X^{(0)}(t'), X^{(0)}(t'') , t' - t''\right) \right\rangle \\  = - \lambda^2  \int \frac{\mathrm{d}k}{2 \pi}\,e^{-D(k^2+r) |t' - t''| - \ell_T^2k^2\left(1 + e^{-\omega_0|t' - t''|}\right)} \frac{k^3\, \sin(2 X_0 k)}{k^2 + r}.
    \label{1p_pert_app_term_2}
\end{multline}
Note that in Eqs.~\eqref{1p_pert_app_term_1} and \eqref{1p_pert_app_term_2}, we kept only the terms of the integrand which are even functions of $k$. In fact, according to Eq.~\eqref{app_exp_avg1}, the translationally-invariant terms in Eqs.~\eqref{eq:app_pointlike_F} and  \eqref{eq:app_pointlike_G} will give rise to additional terms which are odd functions of $k$, which vanish after integrating over the wavevectors.  
Summing Eqs.~\eqref{1p_pert_app_term_1} and  \eqref{1p_pert_app_term_2} yields 
\begin{align}
    &\langle \lambda^2 X^{(2)}(t) \rangle  \label{eq:app_1p_almost_there}  \\ &=  \frac{\lambda^2 }{\kappa} \int\frac{\mathrm{d}k}{2 \pi}\frac{k \sin (2 X_0 k)}{k^2 + r}\int_0^{\infty}\mathrm{d}u \left[D(k^2 + r) - \ell_T^2k^2\omega_0 e^{-\omega_0u} \right] e^{-D(k^2+r)u - \ell_T^2 k^2\left(1 + e^{-\omega_0u}\right)}.  \n 
\end{align}
Note that the integrand of this function can actually be written as a total derivative, i.e., 
\begin{align}
    \left[D(k^2 + r) - \ell_T^2k^2\omega_0 e^{-\omega_0u} \right]&e^{-D(k^2+r)u - \ell_T^2k^2\left(1 + e^{-\omega_0u}\right)} \n \\  &= - \frac{\mathrm{d}}{\mathrm{d}u}\left[e^{-D(k^2+r)u - \ell_T^2k^2\left(1 + e^{-\omega_0u}\right)}\right]. \label{eq:app_F+_der}
\end{align}
Using this fact in Eq.~\eqref{eq:app_1p_almost_there} makes the integral over $u$ straightforward and leads to
\begin{align}
\lambda^2 &\langle  X^{(2)}(t) \rangle \n 
= \frac{\lambda^2 }{\kappa} \int\frac{\mathrm{d}k}{2 \pi}\frac{e^{-2\ell_T^2k^2}k \sin (2 X_0 k)}{k^2 + r}  \n 
\\ &= \frac{\lambda^2 }{4 \kappa} e^{-\frac{2 X_0}{\xi} + \frac{2 \ell_T^2}{ \xi^2}}\left[ \erfc\left( \frac{\sqrt{2}\ell_T}{ \xi} - \frac{X_0}{\sqrt{2 }\ell_T}\right) - \erfc\left( \frac{\sqrt{2}\ell_T}{ \xi} + \frac{X_0}{\sqrt{2}\ell_T}\right) \right] \n   
\\ & \qquad \qquad \qquad \qquad \qquad \qquad \qquad \qquad \qquad    \xrightarrow[{X_0}/{\ell_T} \to \infty]{} \frac{\lambda^2 }{2 \kappa} e^{-\frac{2 X_0}{\xi} + \frac{2 \ell_T^2}{\xi^2}}
\end{align}
where $\erfc(x) =2/\sqrt{\pi}\int_x^{\infty}\mathrm{d}u \, e^{-u^2}$ is the complementary error function. The calculation recovers the result obtained using the Gibbs-Boltzmann distribution reported in~Eq.~(\ref{1p_eq_1d}).

\subsection{The correlation function}
\label{corr_app}
 The lowest-order correction to the particle two-point correlation function is
 \begin{equation}
 \lambda^2 C^{(2)}(t) =    \lambda^2 \langle X^{(2)}(t_1)  X^{(0)}(t_2) + X^{(1)}(t_1)  X^{(1)}(t_2) + X^{(0)}(t_1)  X^{(2)}(t_2)\rangle,
 \end{equation}
 where we denoted $t = t_2 - t_1>0$ and used the stationarity of the process. Then, the correction to the connected  two-point function is given by 
 \begin{align}
 \lambda^2 C_c^{(2)}(t) &= \lambda^2C^{(2)}(t) - 2X_0\lambda^2 \langle X^{(2)} \rangle \n   \\   &=   \lambda^2 \langle X^{(1)}(t_1)  X^{(1)}(t_2) + \tilde{X}(t_1)  X^{(2)}(t_2) + X^{(2)}(t_1)  \tilde{X}(t_2) \rangle,
 \label{C_c_def_app}
 \end{align}
 where $\tilde X(t)$ is the Ornstein-Uhlenbeck process  introduced in \cref{X_0} characterized by the transition probability  in Eq.~\eqref{OU_pdf}.
 We now evaluate each term in the last line in \cref{C_c_def_app}, starting from the first one:
 \begin{multline}
     \lambda^2 \left \langle  X^{(1)}(t_1) X^{(1)}(t_2) \right \rangle \\ = \frac{1}{\gamma_0^2}\int_{-\infty}^{t_1}\!\!\mathrm{d}t_1' \, e^{-\omega_0(t_1 - t_1')} \int_{-\infty}^{t_2}\!\!\mathrm{d}t_2'\, e^{-\omega_0(t_2 - t_2')}\left \langle \Xi\left(X^{(0)}(t_1'), t_1'\right)\Xi\left(X^{(0)}(t_2'), t_2'\right) \right \rangle \\  =  \frac{T}{\gamma_0^2}\int_{-\infty}^{t_1}\!\!\mathrm{d}t_1' \, e^{-\omega_0(t_1 - t_1')} \int_{-\infty}^{t_2}\!\!\mathrm{d}t_2'\, e^{-\omega_0(t_2 - t_2')}\left \langle G\left(X^{(0)}(t_1'),X^{(0)}(t_2'), t_1' - t_2'\right) \right \rangle ,
     \label{bbbbbb}
 \end{multline}
 where we  used Eqs.~\eqref{X_0}  and \cref{eq:eff_dyn_noise} for the correlation of the field-induced noise. 
Now, applying  Eq.~(\ref{app_exp_avg1}) to the expression for $G$ in \cref{eq:app_pointlike_G}, we find
 \begin{align}
    &\left \langle  G\left(X^{(0)}(t'), X^{(0)}(t'') , t' - t''\right) \right\rangle \n \\  
    &\qquad \qquad=   \lambda^2  \int \frac{\mathrm{d}k}{2 \pi}e^{-D(k^2+r) |t' - t''| -\ell_T^2k^2\left(1 - e^{-\omega_0|t' - t''|} \right)} \frac{k^2  }{k^2 + r} \n \\ &\qquad \qquad +    \lambda^2  \int \frac{\mathrm{d}k}{2 \pi}e^{-D(k^2+r) |t' - t''| -\ell_T^2k^2\left(1 + e^{-\omega_0|t' - t''|} \right) }\frac{k^2 \cos(2X_0 k) }{k^2 + r},
\end{align}
where we kept only terms which are even functions of $k$. 
 In order to simplify  the notation, we introduce
 \begin{equation}
     \mathrm{F}(u) =  \int \frac{\mathrm{d}k}{2 \pi}e^{-D(k^2+r)u}\frac{k^2 }{k^2 + r} \left[ e^{-\ell_T^2k^2\left(1 - e^{-\omega_0u} \right)} + e^{-\ell_T^2k^2\left(1 + e^{-\omega_0u} \right)} \cos(2X_0 k) \right]
     \label{dimfull_memo}
 \end{equation}
 which allows us to cast \cref{bbbbbb}, after a suitable change of variables of integration, in the form
 \begin{equation}
     \lambda^2 \left \langle X^{(1)}(t_1) X^{(1)}(t_2) \right \rangle = \frac{\lambda^2 T}{\gamma_0^2}  e^{- \omega_0 t} \int_0^{\infty}\!\!\mathrm{d}t_1'\,e^{-\omega_0 t_1'}\int_{-t}^{\infty}\!\!\mathrm{d}t_2'\, e^{-\omega_0 t_2'} \mathrm{F}(|t_1' - t_2'|).
 \end{equation}
A direct calculation shows that
\begin{align}
    \int_0^{\infty}\!\!\mathrm{d}t_1'\, & e^{-\omega_0 t_1'} \int_{-t}^{\infty}\!\!\mathrm{d}t_2'\, e^{-\omega_0 t_2'} \mathrm{F}(|t_1' - t_2'|)  =  \n \\ &\frac{1}{2 \omega_0 }\left[ \int_0^{\infty}\!\!\mathrm{d}u\, e^{- \omega_0 u} \mathrm{F}(u)  + \int_0^{t}\!\!\mathrm{d}u\, e^{\omega_0 u }\mathrm{F}(u) + e^{2 \omega_0t} \int_t^{\infty}\!\!\mathrm{d}u\, e^{-\omega_0 u }\mathrm{F}(u)  \right]. 
\end{align}
This leads to
\begin{multline}
    \lambda^2 \left \langle  X^{(1)}(t_1)X^{(1)}(t_2) \right \rangle = \\ 
    \frac{\lambda^2 \ell_T^2}{2 \kappa} \omega_0  \left[ \int_0^{\infty}\!\!\mathrm{d}u\, e^{- \omega_0 (u+t)} \mathrm{F}(u)  +  \int_0^{t}\!\!\mathrm{d}u\, e^{-\omega_0(t- u) }\mathrm{F}(u) + \int_t^{\infty}\!\!\mathrm{d}u\, e^{\omega_0 (t-u) }\mathrm{F}(u) \right].
    \label{C_c_1_app}
\end{multline}
We now turn our attention to the second term in the last line of \cref{C_c_def_app}, i.e., 
\begin{align}
    \lambda^2 & \left \langle \tilde X(t_1) X^{(2)}(t_2) \right \rangle \n \\  &=       \frac{1}{\gamma_0^2}\int_{-\infty}^{t_2}\!\!\mathrm{d}t_2'\,\int_{-\infty}^{t_2}\!\!\mathrm{d}t_2''\, e^{-\omega_0 (t_2  - t_2'')}\left \langle \tilde{X}(t_1) \partial_{X^{(0)}(t_2')}\Xi\left(X^{(0)}(t_2'), t_2'\right) \Xi \left( X^{(0)}(t_2''), t_2'' \right) \right \rangle \n \\ &+ \frac{1}{\gamma_0}\int_{-\infty}^{t_2}\!\!\mathrm{d}t_2'\, e^{-\omega_0 (t_2 - t_2')}\int_{-\infty}^{t_2}\!\!\mathrm{d}t_2''\,  \left \langle \tilde{X}(t_1) F\left(X^{(0)}(t_2'), X^{(0)}(t_2''), t_2' - t_2'' \right) \right \rangle,
    \label{term_2}
\end{align}
where we used the expression for $\lambda^2 X^{(2)}$ in \cref{X_2}.  We evaluate the average in the first term of Eq.~\eqref{term_2} using the definition of $G$ in Eqs.~\eqref{eq:eff_dyn_noise} and \eqref{eq:app_pointlike_G}, and the property of the Ornstein-Uhlenbeck process in Eq.~\eqref{app_exp_avg2}, concluding that
\begin{align}
    &\left \langle \tilde{X}(t_1) \partial_{X^{(0)}(t_2')}\Xi\left(X^{(0)}(t_2'), t_2'\right) \Xi \left( X^{(0)}(t_2''), t_2'' \right) \right \rangle \n \\=& -i\lambda^2  T\int\frac{\mathrm{d}k}{2\pi}\frac{e^{-D (k^2 + r)(t_2'- t_2'')}}{k^2 + r}k^3  \left\langle \tilde{X}(t_1)  e^{-ik(X^{(0)}(t_2') - X^{(0)}(t_2''))} \right \rangle \n \\ &-i\lambda^2  T\int\frac{\mathrm{d}k}{2\pi}\frac{e^{-D (k^2 + r)(t_2'- t_2'')}}{k^2 + r}k^3   \left\langle \tilde{X}(t_1) e^{-ik(X^{(0)}(t_2') + X^{(0)}(t_2''))} \right \rangle,  \label{term_2_1}   \\ &=-\lambda^2 T \ell_T^2  \int\frac{\mathrm{d}k}{2\pi}e^{-D (k^2 + r)(t_2'- t_2'') -\ell_T^2k^2 \left(1 - e^{-\omega_0(t_2' - t_2'')}\right)}\frac{k^4}{k^2 + r} \left(e^{-\omega_0 |t_1 - t_2'|} - e^{-\omega_0 |t_1 - t_2''|} \right) \n \\ &- \lambda^2 T \ell_T^2 \int\frac{\mathrm{d}k}{2\pi}e^{-D (k^2 + r)(t_2'- t_2'')-\ell_T^2k^2 \left(1 + e^{-\omega_0(t_2' - t_2'')}\right)}\frac{k^4 \cos(2X_0 k) }{k^2 + r} \left(e^{-\omega_0 |t_1 - t_2'|} + e^{-\omega_0 |t_1 - t_2''|} \right). \n
\end{align}
Similarly, we calculate the average in the second term in Eq.~\eqref{term_2}, utilizing Eqs.~\eqref{app_exp_avg2}  and \eqref{eq:app_pointlike_F}. This yields
\begin{align}
    &\left \langle \tilde{X}(t_1) F\left(X^{(0)}(t_2'), X^{(0)}(t_2''), t_2' - t_2'' \right) \right \rangle  \n \\ &= -i\lambda^2  D \int\frac{\mathrm{d}k}{2\pi}\, e^{-D (k^2 + r)(t_2'- t_2'')} k 
     \left\langle \tilde{X}(t_1) e^{-ik(X^{(0)}(t_2') - X^{(0)}(t_2''))} \right \rangle  \n \\ &\quad  \, + i\lambda^2  D \int\frac{\mathrm{d}k}{2\pi}\, e^{-D (k^2 + r)(t_2'- t_2'')} k  \left\langle \tilde{X}(t_1) e^{-ik(X^{(0)}(t_2') + X^{(0)}(t_2''))} \right \rangle \label{term_2_2} \\ &=-\lambda^2  D  \ell_T^2 \int\frac{\mathrm{d}k}{2\pi}\, e^{-D (k^2 + r)(t_2'- t_2'')} k^2 e^{-\frac{T}{\kappa}k^2 \left(1 - e^{-\omega_0(t_2' - t_2'')}\right)} \left(e^{-\omega_0 |t_1 - t_2'|} - e^{-\omega_0 |t_1 - t_2''|} \right)   \n \\ & \quad  \,+ \lambda^2  D \ell_T^2\int\frac{\mathrm{d}k}{2\pi}\, e^{-D (k^2 + r)(t_2'- t_2'')-\ell_T^2k^2 \left(1 + e^{-\omega_0(t_2' - t_2'')}\right)} k^2 \cos(2X_0 k) \left(e^{-\omega_0 |t_1 - t_2'|} + e^{-\omega_0 |t_1 - t_2''|} \right).\n 
\end{align}
Accordingly, using Eqs.~\eqref{term_2_1} and \eqref{term_2_2} in \cref{term_2} renders
\begin{align}
    \lambda^2 & \left \langle \tilde{X}(t_1) X^{(2)}(t_2) \right \rangle \n \\ &=    - \frac{\lambda^2  \ell_T^2}{ \gamma_0}  \int_{-\infty}^{t_2}\!\!\mathrm{d}t_2'\, e^{-\omega_0 (t_2 - t_2')}\int_{-\infty}^{t_2'}\!\!\mathrm{d}t_2''\, \left(e^{-\omega_0 |t_1 - t_2'|} - e^{-\omega_0 |t_1 - t_2''|} \right) \times \n \\ & \qquad  \int\frac{\mathrm{d}k}{2\pi}\, e^{-D (k^2 + r)(t_2'- t_2'') - \ell_T^2\left(1 - e^{-\omega_0(t_2' - t_2'') }\right)}\frac{k^2}{k^2 + r}  \left[ D(k^2 + r) + \ell_T^2k^2 \omega_0e^{-\omega_0 (t_2' - t_2'')} \right]  \n \\   &+ \frac{\lambda^2  \ell_T^2}{ \gamma_0} \int_{-\infty}^{t_2}\!\!\mathrm{d}t_2'\,e^{-\omega_0 (t_2 - t_2')}\int_{-\infty}^{t_2'}\!\!\mathrm{d}t_2'' \, \left(e^{-\omega_0 |t_1 - t_2'|} + e^{-\omega_0 |t_1 - t_2''|} \right) \times  \label{term_2_almost}  \\ & \qquad\int\frac{\mathrm{d}k}{2\pi}e^{-D (k^2 + r)(t_2'- t_2'') - \ell_T^2\left(1 + e^{-\omega_0(t_2' - t_2'')} \right) }\frac{k^2 \cos(2x_0k)}{k^2 + r}   \left[ D(k^2 + r) - \ell_T^2k^2 \omega_0e^{-\omega_0 (t_2' - t_2'')} \right]. \n
\end{align}
We decompose $\mathrm{F}$ defined in \cref{dimfull_memo} into $\mathrm{F}(u) = \mathrm{F}^+(u) + \mathrm{F}^-(u)$ with 
\begin{align}
     &\mathrm{F}^+(u) =  \int \frac{\mathrm{d}k}{2 \pi}\, e^{-D(k^2+r)u -\ell_T^2k^2\left(1 - e^{-\omega_0u} \right)}\frac{k^2 }{k^2 + r},  
     \label{F+_dimfull_app} \\ 
     &\mathrm{F}^-(u) =  \int \frac{\mathrm{d}k}{2 \pi}\, e^{-D(k^2+r)u -\ell_T^2k^2\left(1 + e^{-\omega_0u} \right)}\frac{k^2 \cos(2X_0 k)}{k^2 + r},  
     \label{F-_dimfull_app}
 \end{align}
 and, similarly to Eq.~\eqref{eq:app_F+_der}, we note  that
\begin{align}  
\left[ D (k^2 + r) + \ell_T^2k^2 \omega_0e^{-\omega_0u} \right]&e^{-D(k^2+r)u -\ell_T^2k^2\left(1 + e^{-\omega_0u} \right)}  \n \\  &= - \frac{\mathrm{d}}{\mathrm{d}u}\left[e^{-D(k^2+r)u - \ell_T^2k^2\left(1 - e^{-\omega_0u}\right)}\right]
     \label{eq:app_F-_der}
\end{align}
 We can now change the variables of integration in Eq.~\eqref{term_2_almost} to $u = t_2' - t_2''$ and $v = t_1 - t_2'$, apply Eqs.~\eqref{eq:app_F+_der} and \eqref{eq:app_F-_der} along with the definitions of $\rm F^+(u)$ and $ \rm F^-(u)$ in Eqs.~\eqref{F+_dimfull_app} and \eqref{F-_dimfull_app}, respectively, to \cref{term_2_almost}, and write
\begin{align}
   \lambda^2 &\left \langle \tilde{X}(t_1) X^{(2)}(t_2) \right \rangle \n  \\ &=  \frac{\lambda^2  \ell_T^2}{ \gamma_0}e^{-\omega_0 t}\int_{-t}^{\infty}\!\!\mathrm{d}v\,e^{-\omega_0 v}\int_0^{\infty}\!\!\mathrm{d}u\,  \left(e^{-\omega_0 |v|} - e^{-\omega_0 |v+ u|} \right)\frac{\mathrm{d}\mathrm{F}^+(u)}{\mathrm{d}u}   \n \\ & - \frac{\lambda^2  \ell_T^2}{ \gamma_0}e^{-\omega_0 t}\int_{-t}^{\infty}\!\!\mathrm{d}v\, e^{-\omega_0 v}\int_0^{\infty}\!\!\mathrm{d}u\,   \left(e^{-\omega_0 |v|} + e^{-\omega_0 |v + u|} \right)\frac{\mathrm{d}\mathrm{F}^-(u)}{\mathrm{d}u}, \label{cccccccc}  \\ &=
    \frac{\lambda^2  \ell_T^2}{ \gamma_0}e^{-\omega_0 t} 
    \left[ 2\mathrm{F}^-(0)\int_{-t}^\infty\!\! \mathrm{d}v\, e^{-\omega_0 (v + |v|)} + 
    \int_{-t}^{\infty}\!\!\mathrm{d}v\,e^{-\omega_0v}\int_0^{\infty}\!\!\mathrm{d}u\,\mathrm{F}(u)\frac{\mathrm{d}}{\mathrm{d}u}e^{-\omega_0|u + v|} \right]. \n
\end{align}
Using the fact that
\begin{multline}
    \int_{-t}^{\infty}\!\!\mathrm{d}v\, e^{-\omega_0v}\int_0^{\infty}\!\!\mathrm{d}u\,\mathrm{F}(u)\frac{\mathrm{d}}{\mathrm{d}u}e^{-\omega_0|u + v|} = \\  \frac{1}{2}\left\{ \int_0^t\mathrm{d}u\,\left[2\omega_0(t - u)-1\right]e^{\omega_0 u} \mathrm{F}(u) - e^{2\omega_0t}\int_t^{\infty}\!\!\mathrm{d}u\,e^{-\omega_0 u}\mathrm{F}(u) \right\},
\end{multline}
one can cast Eq.~\eqref{cccccccc} in the form
\begin{align}
    \lambda^2&\left \langle \tilde{X}(t_1) X^{(2)}(t_2)  \right \rangle \qquad\qquad\qquad  
    \n \\ & \qquad =\frac{\lambda^2  \ell_T^2}{ \kappa } \left[ e^{-\omega_0t}\mathrm{F}^-(0)\left(1 + 2 \omega_0t\right)  + \omega_0 \int_0^t \mathrm{d}u \, \omega_0(t-u) e^{-\omega_0(t-u)}\mathrm{F}(u)\right] \label{C_c_2_app} 
    \\ & \qquad  -\frac{\lambda^2  \ell_T^2 }{2 \kappa}\omega_0\left[ \int_0^t \mathrm{d}u\, e^{-\omega_0(t-u)}\mathrm{F}(u) +\int_t^{\infty}\!\!\mathrm{d}u\,e^{\omega_0(t-u)}\mathrm{F}(u) \right].\n
\end{align}
Finally, the calculation of the third term in Eq.~\eqref{C_c_def_app} is very similar to that of the second one therein. Accordingly, we simply report the final expression  
\begin{align}
    &\lambda^2\left \langle  X^{(2)}(t_1) \tilde{X}(t_2) \right \rangle 
    = \frac{\lambda^2  \ell_T^2}{2 \kappa}\left[2e^{-\omega_0 t} \mathrm{F}^-(0)  - \omega_0 \int_0^{\infty}\!\!\mathrm{d}u\,e^{-\omega_0(u+t)}\mathrm{F}(u) \right].
    \label{C_c_3_app}
\end{align}
 Summing the results of Eqs.~\eqref{C_c_1_app}, \eqref{C_c_2_app}, and  \eqref{C_c_3_app} allows us to cast the expression for the connected part of the two-point function in Eq.~(\ref{C_c_def_app}) in the form
\begin{equation}
    \lambda^2C_c^{(2)}(t) =  \frac{\lambda^2 \ell_T^2 }{\kappa}\left[2 e^{-\omega_0t}\mathrm{F}^-(0)(1 + \omega_0 t) + \omega_0 \int_0^t\mathrm{d}u \, \omega_0(t - u)e^{-\omega_0(t-u)}\mathrm{F}(u) \right],
    \label{eq:C_c_app_final}
\end{equation}
which is the expression provided in Eq.~\eqref{C_c_mem}, which is expressed in terms of the dimensionless memory defined below in \cref{dimless_memo_app}.
For the sake of completeness, we note that a straightforward calculation renders the complete expression of $\mathrm{F}^+(u)$ introduced in Eq.~\eqref{F+_dimfull_app} in the form
\begin{multline}
    \mathrm{F}^+(u) = \frac{e^{-\frac{Du}{\xi^2}}}{\sqrt{4 \pi (Du + \ell_T^2(1 - e^{-\omega_0u}))}} - \frac{1}{2\xi}e^{\frac{\ell_T^2}{ \xi^2}(1 - e^{-\omega_0u})} \erfc\left(\sqrt{\frac{Du}{\xi^2} + \frac{\ell_T^2}{\xi^2}(1 - e^{-\omega_0u})}\right)   \\ \xrightarrow[\xi \to \infty]{} \frac{1}{\sqrt{4 \pi (Du + \ell_T^2(1 - e^{-\omega_0u}))}}=\mathrm{F}_c^+(u).
    \label{F+_app}
\end{multline}
The limit $\xi \to \infty$ of the expression above is reported in the main text in Eq.~\eqref{memo_crit_pl}. In addition, the complete expression of $\mathrm{F}^-(u)$ can be obtained by performing the integral in Eq.~\eqref{F-_dimfull_app}:
\begin{multline}
    \mathrm{F}^-(u) =  \frac{e^{-\left(\frac{Du}{\xi^2} + \frac{X_0^2}{Du + \ell_T^2(1 + e^{-\omega_0u})} \right)}}{\sqrt{4 \pi (Du + \ell_T^2(1 + e^{-\omega_0u}))}} \\ - \frac{1}{4\xi}e^{-\frac{2X_0}{\xi}+\frac{\ell_T^2}{\xi^2}(1 + e^{-\omega_0u})} \erfc\left(\sqrt{\frac{Du}{\xi^2} + \frac{\ell_T^2}{\xi^2}(1 + e^{-\omega_0u})} - \frac{X_0}{\sqrt{Du + \ell_T^2(1 + e^{-\omega_0u})}}\right)  \\ - \frac{1}{4\xi}e^{\frac{2X_0}{\xi} +\frac{\ell_T^2}{ \xi^2}(1 + e^{-\omega_0u})} \erfc\left(\sqrt{\frac{Du}{\xi^2} + \frac{\ell_T^2}{\xi^2}(1 + e^{-\omega_0u})} + \frac{X_0}{\sqrt{Du + \ell_T^2(1 + e^{-\omega_0u})}}\right)     \\ \xrightarrow[\xi \to \infty]{} \frac{e^{-\frac{X_0^2}{Du + \ell_T^2(1 + e^{-\omega_0u})}}}{\sqrt{4 \pi (Du + \ell_T^2(1 + e^{-\omega_0u}))}} = \mathrm{F}_c^-(u).
    \label{F-_app}
\end{multline}
The limit $\xi \to \infty$ of the expression above is given  in Eq.~\eqref{memo_crit_min} of the main text.
Finally, we note from Eq.~\eqref{F-_app} that
\begin{align}
    \mathrm{F}^-(0) =  \frac{e^{- \frac{X_0^2}{ 2\ell_T^2} }}{\sqrt{8 \pi  \ell_T^2}}  &- \frac{1}{4\xi}e^{-\frac{2X_0}{\xi}+\frac{2\ell_T^2}{\xi^2}} \erfc\left(\frac{\sqrt{2}\ell_T}{\xi} - \frac{X_0}{\sqrt{2}\ell_T}\right) \n \\ &- \frac{1}{4\xi}e^{\frac{2X_0}{\xi} +\frac{\ell_T^2}{ \xi^2}(1 + e^{-\omega_0u})} \erfc\left(\frac{\sqrt{2}\ell_T}{\xi} +  \frac{X_0}{\sqrt{2}\ell_T}\right) \n    \\ &\xrightarrow[X_0/\ell_T \to \infty]{} - \frac{1}{2\xi}e^{-\frac{2X_0}{\xi}+\frac{2\ell_T^2}{\xi^2}}.
    \label{F-_0_app}
\end{align}
Lastly, introducing the dimensionless memory $\mathcal{F}$ by
\begin{equation}
    \frac{1}{\xi}\mathcal{F}(\omega_0t) =  \mathrm{F}(t)
    \label{dimless_memo_app}
\end{equation}
and combining  the results in Eqs.~\eqref{eq:C_c_app_final} and \eqref{F-_0_app}, renders the final expression of $\lambda^2C_c^{(2)}(t)$ reported in Eqs.~\eqref{C_c_2}, \eqref{C_c_C}, and \eqref{C_c_mem} of the main text.

\section{Power Spectral Density}
\label{PSD_app}

In order to derive the power spectral density reported in Eq.~\eqref{PSD}, we use
\begin{align}
    \int_{\mathbb{R}}\mathrm{d}t\, e^{i \omega t} \int_{0}^{\omega_0 |t|}\mathrm{d}u \,& (\omega_0 |t| - u)e^{-(\omega_0 |t| - u)}\mathcal{F}(u) \n  \\  
     &= \int_{0}^{\infty}\!\!\mathrm{d}t\,e^{i \omega t} \int_{0}^{\omega_0 t}\mathrm{d}u\,  (\omega_0 t - u)e^{-(\omega_0 t - u)}\mathcal{F}(u)  + \mathrm{c.c.} \n  \\ &= \frac{1}{\omega_0}\int_0^{\infty}\!\!\mathrm{d}u\, e^{i\frac{\omega}{\omega_0}u}\mathcal{F}(u)\int_u^{\infty}\!\!\mathrm{d}t\, e^{-(t-u)\left(1 - i\frac{\omega}{\omega_0}\right)}(t-u)  + \mathrm{c.c. } \n \\ & \quad \frac{1}{\omega_0}\frac{\omega_0^2}{(\omega_0 - i \omega)^2}\int_0^{\infty}\mathrm{d}ue^{i\frac{\omega}{\omega_0}u}\mathcal{F}(u) + \mathrm{c.c.},  
\end{align}
where $\rm c.c.$ denotes the complex conjugate. Then, using the properties of trigonometric functions, yields
\begin{equation}
    \frac{\omega_0^2}{(\omega_0 - i \omega)^2} e^{i\frac{\omega}{\omega_0}u}  + \mathrm{c.c.}  = \frac{2\omega_0^4}{(\omega_0^2 +  \omega^2)^2} \left[\left(1 -  \frac{\omega^2}{\omega_0^2}\right) \cos\left(u \frac{\omega}{\omega_0} \right)  -2\frac{\omega}{\omega_0}\sin\left(u \frac{\omega}{\omega_0} \right)  \right],
\end{equation}
which proves the form of the power spectral density given in Eq.~(\ref{PSD}).  
\section*{References}
\providecommand{\newblock}{}

\end{document}